\documentclass[sn-vancouver,Numbered]{sn-jnl}

\usepackage{graphicx}
\usepackage{multirow}
\usepackage{amsmath,amssymb,amsfonts}
\usepackage{amsthm}
\usepackage{mathrsfs}
\usepackage[title]{appendix}
\usepackage{xcolor}
\usepackage{textcomp}
\usepackage{manyfoot}
\usepackage{booktabs}
\usepackage{algorithm}
\usepackage{algorithmicx}
\usepackage{algpseudocode}
\usepackage{listings}
\usepackage{caption}
\usepackage{subcaption}

\renewcommand{\dot}{\boldsymbol{\cdot}}
\renewcommand{\vec}[1]{\boldsymbol{#1}}
\newcommand{\unitvec}[1]{\hat{\boldsymbol{#1}}}

\algnewcommand{\IIf}[1]{\State\algorithmicif\ #1\ \algorithmicthen}
\algnewcommand{\EndIIf}{\unskip\ \algorithmicend\ \algorithmicif}

\theoremstyle{thmstyleone}%
\theoremstyle{thmstyletwo}%

\theoremstyle{thmstylethree}%

\begin{document}
\title[CBRM on the CPC]{Concatenated Backward Ray Mapping on the Compound Parabolic Concentrator}

\author*[1]{\fnm{Willem} \sur{Jansen}}\email{w.g.t.jansen@tue.nl}
\author[1]{\fnm{Martijn} \sur{Anthonissen}}\email{m.j.h.anthonissen@tue.nl}\equalcont{These authors contributed equally to this work.}
\author[1]{\fnm{Jan} \sur{ten Thije Boonkkamp}}\email{j.h.m.tenthijeboonkkamp@tue.nl}\equalcont{These authors contributed equally to this work.}
\author[2,1]{\fnm{Wilbert} \sur{IJzerman}}\email{wilbert.ijzerman@signify.com}\equalcont{These authors contributed equally to this work.}


\affil*[1]{\centering Department of Mathematics and Computer Science,\\ Eindhoven University of Technology, PO Box 513, 5600 MB Eindhoven,\\ The Netherlands}
\affil[2]{\centering Signify Research, High Tech Campus 7, 5656 AE Eindhoven,\\ The Netherlands}

\abstract{Concatenated backward ray mapping is an alternative for ray tracing in 2D. It is based on the phase-space description of an optical system. Phase space is the set of position and direction coordinates of light rays intersecting a surface. The original algorithm \cite{Filosa2021} is limited to optical systems consisting of only straight surfaces; we generalize it to accommodate curved surfaces. The algorithm is applied to a standard optical system, the compound parabolic concentrator. We compare the accuracy and speed of the generalized algorithm, the original algorithm and Monte Carlo ray tracing. The results show that the generalized algorithm outperforms both other methods.}


\maketitle

\section{Introduction}\label{sec:1}
The illumination optics industry deals with the design of optical systems. An optical system consists of a light source, optical components such as lenses and reflectors and a target which is either a receiver or an aperture. Light emitted at the source of the optical system propagates through the system and forms an intensity distribution at the target. The shape of the target distribution depends on the optical system and the intensity distribution at the source. The goal in illumination optics is to obtain a desired intensity distribution at the target. Designing an optical system is an iterative process during which the target distribution is computed many times. Therefore, there is a need for fast and accurate simulation methods.

The target distribution is typically computed using Monte Carlo (MC) or Quasi-Monte Carlo (QMC) ray tracing. MC ray tracing is based on a probabilistic interpretation of the source distribution \cite{Jensen2003}. Many randomly distributed rays are traced from source to target; the intensity distribution is found by dividing the target into equal cells and counting the number of rays in each cell. QMC ray tracing was introduced as a faster alternative to MC ray tracing. The difference between them is that QMC ray tracing distributes the rays along low discrepancy sequences \cite{caflisch_1998}. MC ray tracing is a slow and expensive procedure with a convergence rate of $\mathcal{O}(1 / \sqrt{\text{Nr}})$ \cite{FilosaThesis}, where Nr is the number of rays traced. QMC ray tracing performs better with a convergence rate of $\mathcal{O}(1 / \text{Nr})$ \cite{FilosaThesis}, but it is still an expensive procedure.

Concatenated backward ray mapping (CBRM) is an alternative to (Q)MC ray tracing in 2D that uses the phase space (PS) \cite{Filosa2021}. The PS of a surface is defined by the position and direction coordinates of all rays that interact with the surface. An optical surface in 2D is a line segment or a curved segment, but we will still refer to it as a surface. The algorithm determines which light rays emitted by the source reach the target at a certain angle by tracing backward; only those rays are then traced from source to target, resulting in a significant reduction in the number of rays needed. Doing this for all angles at the target gives the intensity distribution. Numerical results show that CBRM computes the intensity distribution more accurately and with less computation time than (Q)MC ray tracing \cite{Filosa2021}. However, CBRM requires the optical system to consist of straight surfaces. We generalize CBRM to accommodate curved optical surfaces.

\begin{figure}[htb]
    \begin{minipage}[t]{0.45\textwidth}
        \centering
        \includegraphics[width=\textwidth]{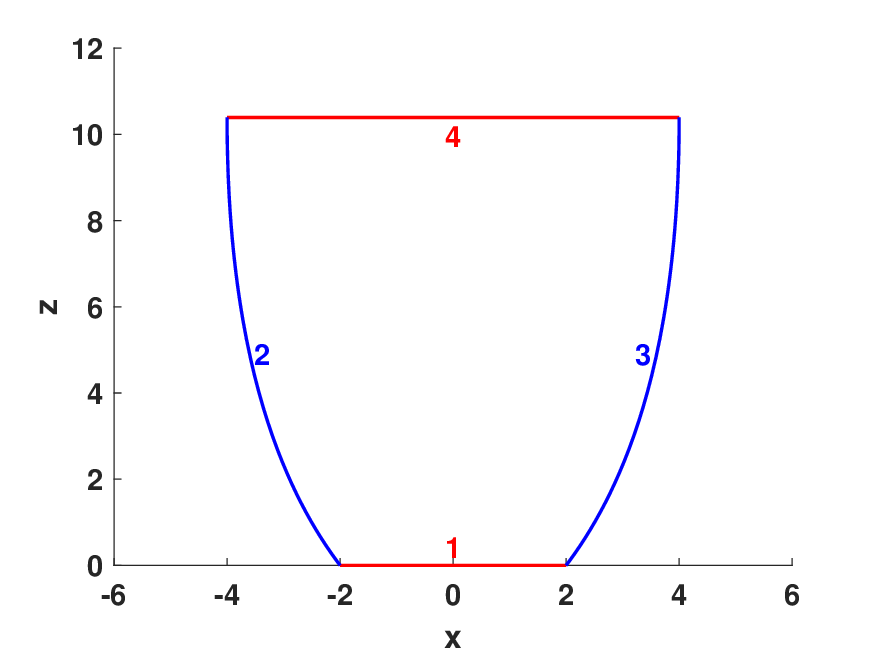}
        \caption{The CPC with source (1), reflectors (2, 3), and target (4).}
        \label{fig:cpc}
    \end{minipage}
    \hspace{0.1\textwidth}
    \begin{minipage}[t]{0.45\textwidth}
        \centering
        \includegraphics[width=\textwidth]{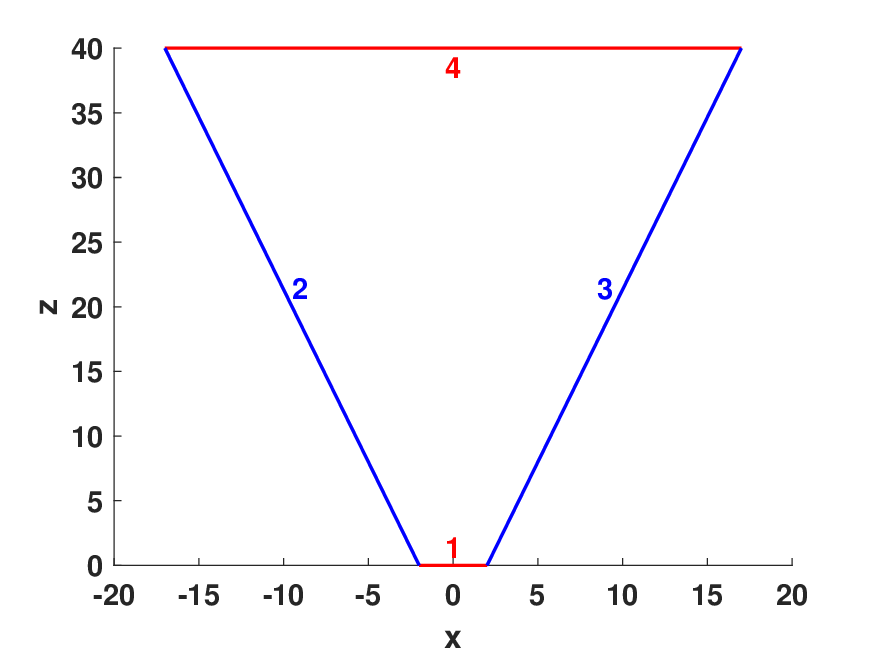}
        \caption{The two-faceted cup with source (1), reflectors (2, 3), and target (4).}
        \label{fig:cup}
    \end{minipage}
\end{figure}

The purpose of this paper is to  introduce the generalized CBRM algorithm and apply it to the compound parabolic concentrator (CPC) \cite{chaves2008}. The CPC (Fig. \ref{fig:cpc}) is a standard optical system which collects light from a Lambertian source and reshapes it to a focused beam. Before we introduce the generalized CBRM algorithm we first explain CBRM and apply it to the two-faceted cup (Fig. \ref{fig:cup}). The performance of CBRM and generalized CBRM is compared on the CPC. Since CBRM requires the optical system to consist of straight surfaces, it is applied to a discretized CPC.

The structure of this paper is as follows. Section \ref{sec:2} defines the phase space of an optical surface. In Section \ref{sec:3} we describe the concatenated backward ray mapping algorithm. Section \ref{sec:4} explains the generalized algorithm. Section \ref{sec:5} describes the numerical experiments that compare the algorithms on the (discretized) CPC. Section \ref{sec:6} shows the results of the experiments. In Section \ref{sec:7} we draw conclusions.

\section{Phase Space}\label{sec:2}
Concatenated backward ray mapping is based on the phase-space (PS) \cite{FilosaThesis} description of an optical system. Phase space is the set of position and direction coordinates of rays intersecting a surface. It is a two-dimensional space for 2D surfaces. The position coordinate $q \in Q$ is the $x$-coordinate of the intersection of the ray with the surface. The direction coordinate $p \in P$ is given by $p = n \sin{\theta}$, where $\theta \in [-\pi/2, \pi/2]$ is the angle between the ray and the inward facing unit surface normal $\unitvec{\nu}$ and $n$ is the index of refraction. We consider only optical systems formed by reflective surfaces, therefore refraction is not taken into account and $n=1$. The index $n$ will be omitted from now on. PS is indicated with $S = Q \times P$. Every optical surface has a source PS and a target PS. Source PS describes light emitted by the surface and target PS describes light reaching the surface. The light source only emits light, so it only has a source PS; the target only receives light, so it only has a target PS.

The optical surfaces of the system are numbered $1, \dots, \text{N}$ where 1 is the light source and $\text{N}$ is the target. The source and target PS of surface $j$ are indicated with $\text{S}_j$ and $\text{T}_j$, respectively. The coordinates of a ray reaching surface $j$ is indicated with $(q_{\text{t}, j}, p_{\text{t}, j}) \in \text{T}_j$. After reflection, the ray is emitted by the surface from the same position but with a new direction. The ray now has the coordinates $(q_{\text{s}, j}, p_{\text{s}, j}) \in \text{S}_j$. Note that $q_{\text{t}, j} = q_{\text{s}, j}$ while $p_{\text{s}, j}$ is obtained by applying the law of reflection to the ray.

The phase spaces $\text{S}_j$ and $\text{T}_j$ are divided into regions $\text{S}_{j,k}$ and $\text{T}_{j,l}$. $\text{S}_{j,k}$ is the region of $\text{S}_j$ containing all light rays emitted by $j$, illuminating surface $k \in \{2,\dots,\text{N}\}$ assuming that $j$ acts as a light source. $\text{T}_{j,l}$ is the region of $\text{T}_j$ containing all light rays emitted by surface $l \in \{1,\dots,\text{N}-1\}$, illuminating surface $j$ assuming that $l$ acts as a light source. Note that $\text{S}_j$ and $\text{T}_j$ may contain empty regions. The boundaries $\partial \text{S}_{j,k}$ are connected to the boundaries $\partial \text{T}_{k,j}$ for every $j \in \{1, \dots, \text{N}-1\}$ and $k \in \{2, \dots, \text{N}\}$ by the edge-ray principle \cite{FilosaThesis}. These boundaries can be determined analytically and are formed by four curves:

\vspace{-7mm}
\begin{equation}\label{eq:1}
    \begin{split}
        \partial \text{S}_{j,k} &= \partial \text{S}^1_{j,k} \cup \partial \text{S}^2_{j,k} \cup \partial \text{S}^3_{j,k} \cup \partial \text{S}^4_{j,k},\\
        \partial \text{T}_{k,j} &= \partial \text{T}^1_{k,j} \cup \partial \text{T}^2_{k,j} \cup \partial \text{T}^3_{k,j} \cup \partial \text{T}^4_{k,j}.
    \end{split}
\end{equation}

\begin{figure}[htb]
    \begin{subfigure}[t]{0.45\textwidth}
        \centering
        \includegraphics[width=\textwidth]{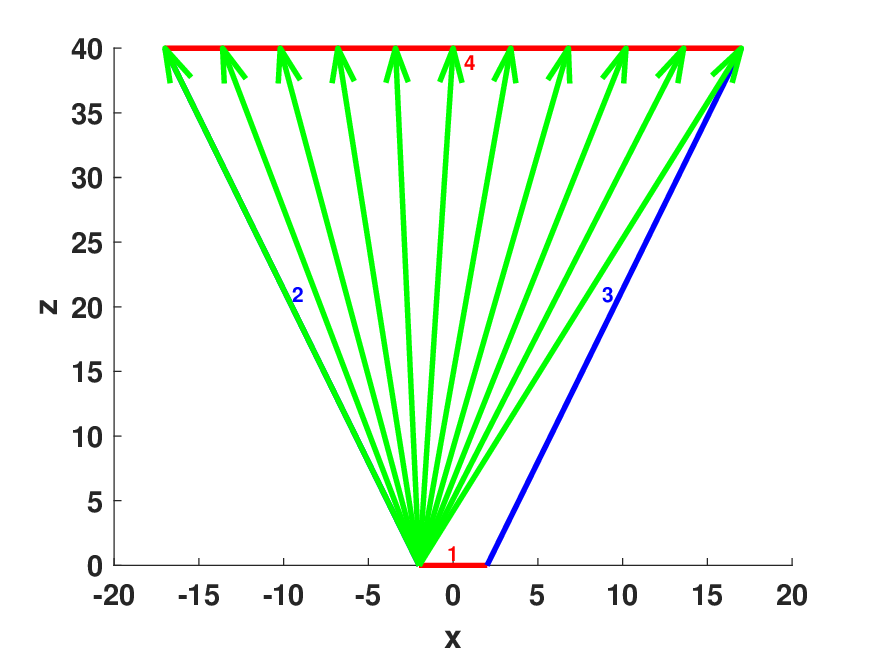}
        \caption{Rays that leave the left endpoint of the source and trace out the target forming $\partial \text{S}^1_{1,4}$ and $\partial \text{T}^1_{4,1}$.}
        \label{fig:boundary_rays_1}
    \end{subfigure}
    \hspace{0.1\textwidth}
    \begin{subfigure}[t]{0.45\textwidth}
        \centering
        \includegraphics[width=\textwidth]{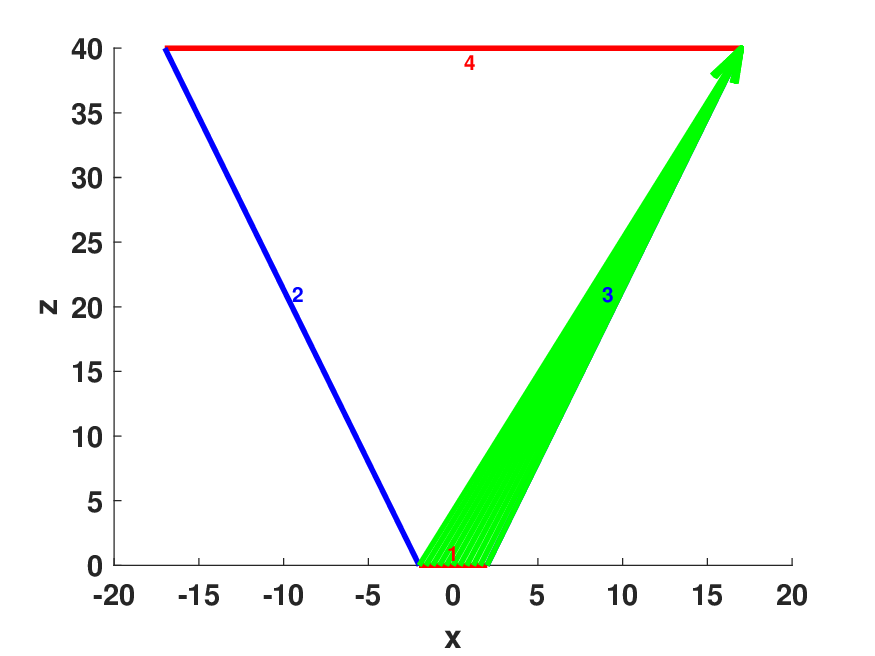}
        \caption{Rays that trace out the source and hit the right endpoint of the target forming $\partial \text{S}^2_{1,4}$ and $\partial \text{T}^2_{4,1}$.}
        \label{fig:boundary_rays_2}
    \end{subfigure}
    \begin{subfigure}[t]{0.45\textwidth}
        \centering
        \includegraphics[width=\textwidth]{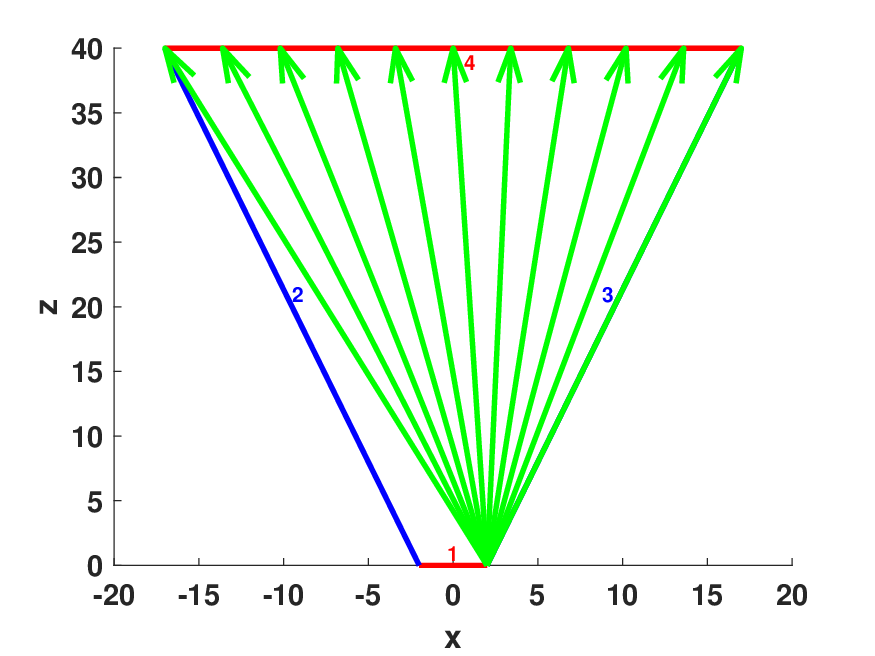}
        \caption{Rays that leave the right endpoint of the source and trace out the target forming $\partial \text{S}^3_{1,4}$ and $\partial \text{T}^3_{4,1}$.}
        \label{fig:boundary_rays_3}
    \end{subfigure}
    \hspace{0.1\textwidth}
    \begin{subfigure}[t]{0.45\textwidth}
        \centering
        \includegraphics[width=\textwidth]{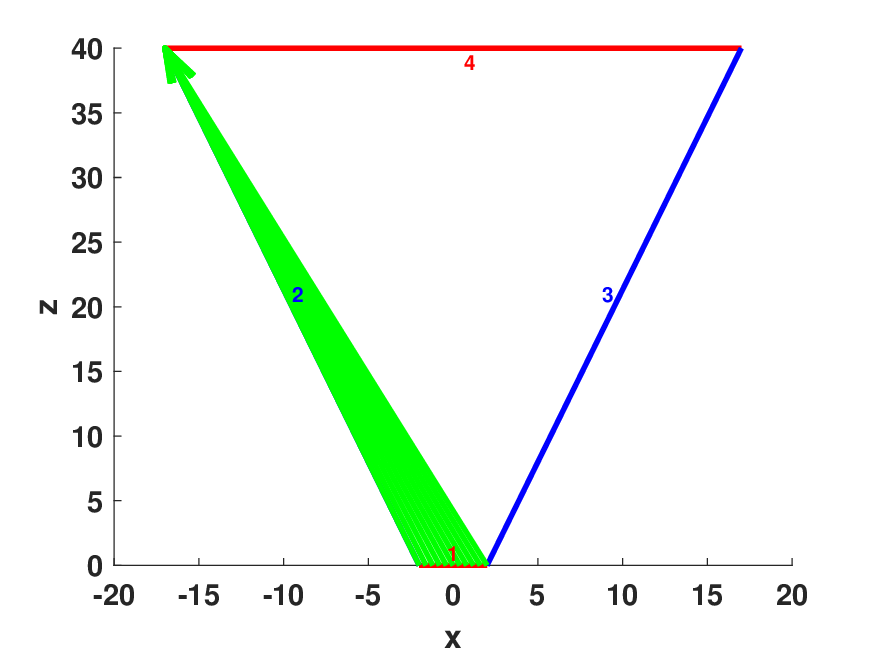}
        \caption{Rays that trace out the source and hit the left endpoint of the target forming $\partial \text{S}^4_{1,4}$ and $\partial \text{T}^4_{4,1}$.}
        \label{fig:boundary_rays_4}
    \end{subfigure}
    \caption{Rays on the boundaries of the regions $\text{S}_{1,4}$ and $\text{T}_{4,1}$ in the two-faceted cup.}
    \label{fig:boundary_rays}
\end{figure}

\begin{figure}[H]
    \begin{minipage}[t]{0.45\textwidth}
        \centering
        \includegraphics[width=\textwidth]{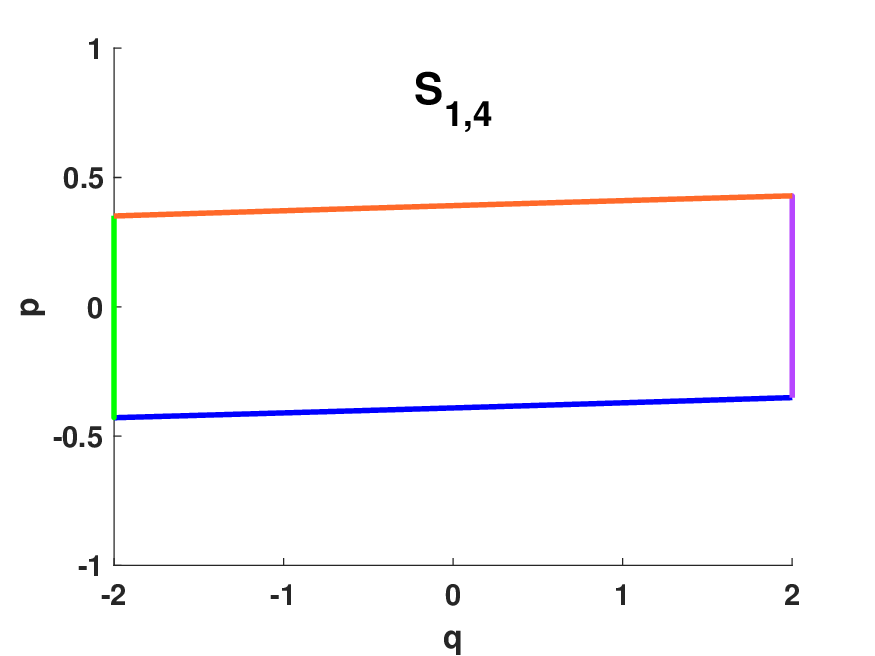}
        \caption{Boundary of the region $\text{S}_{1,4}$ of the two-faceted cup consisting of segments $\partial \text{S}^1_{1,4}$ (green), $\partial \text{S}^2_{1,4}$ (blue), $\partial \text{S}^3_{1,4}$ (purple) and $\partial \text{S}^4_{1,4}$ (orange).}
        \label{fig:source_boundary}
    \end{minipage}
    \hspace{0.1\textwidth}
    \begin{minipage}[t]{0.45\textwidth}
        \centering
        \includegraphics[width=\textwidth]{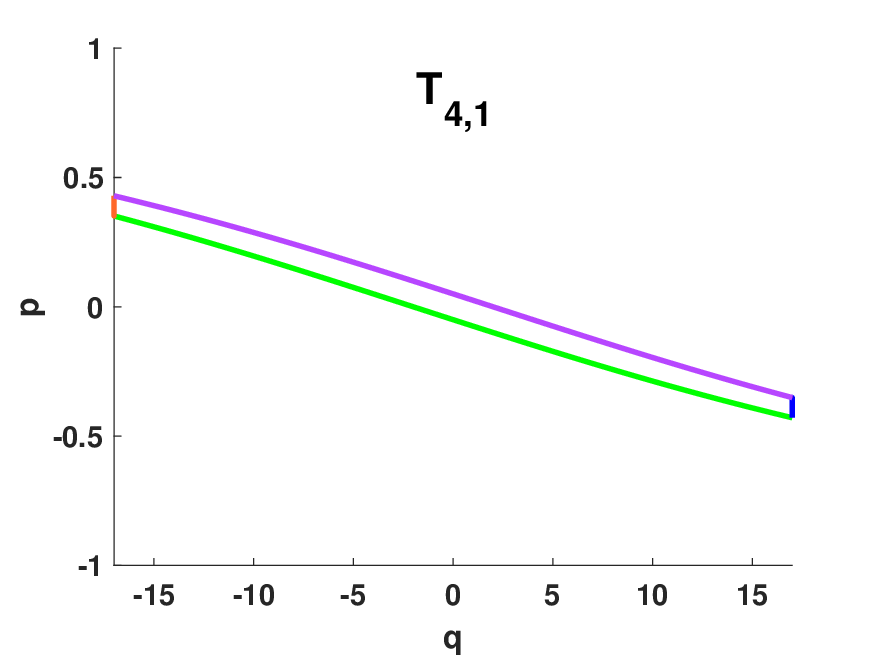}
        \caption{Boundary of the region $\text{T}_{4,1}$ of the two-faceted cup consisting of segments $\partial \text{T}^1_{4,1}$ (green), $\partial \text{T}^2_{4,1}$ (blue), $\partial \text{T}^3_{4,1}$ (purple) and $\partial \text{T}^4_{4,1}$ (orange).}
        \label{fig:target_boundary}
    \end{minipage}
\end{figure}

Given two surfaces $j$ and $k$, $\partial \text{S}_{j,k}$ and $\partial \text{T}_{k,j}$ are determined as follows. $\partial \text{S}^1_{j,k}$ and $\partial \text{T}^1_{k,j}$ are formed by the set of rays originating at the left endpoint of $j$ tracing out surface $k$. $\partial \text{S}^2_{j,k}$ and $\partial \text{T}^2_{k,j}$ are formed by the set of rays tracing out surface $j$ reaching the right endpoint of $k$. $\partial \text{S}^3_{j,k}$ and $\partial \text{T}^3_{k,j}$ are formed by the set of rays originating at the right endpoint of $j$ tracing out surface $k$. $\partial \text{S}^4_{j,k}$ and $\partial \text{T}^4_{k,j}$ are formed by the set of rays tracing out surface $j$ reaching the left endpoint of $k$. As an example Fig. \ref{fig:boundary_rays} shows these sets of rays in the two-faceted cup where $j=1$ and $k=4$, i.e., the source and target. The segments formed by these rays in $\text{S}_{1}$ are shown in Fig. \ref{fig:source_boundary}. The segments formed in $\text{T}_{4}$ are shown in Fig. \ref{fig:target_boundary}.

Previously\cite{Filosa2021,FilosaThesis} we only had analytic expressions of the boundaries of Eq. (\ref{eq:1}) when $j$ and $k$ were straight surfaces. Here, we introduce a general analytic expression for each boundary of Eq. (\ref{eq:1}) when surfaces $j$ and $k$ are described by parametric equations. Let surface $j$ be described by the parameterization $\vec{P}_{j}(\gamma) = \big( x_{j}(\gamma), z_{j}(\gamma) \big) \ (\gamma_{\text{min}} \leq \gamma \leq \gamma_{\text{max}})$ and surface $k$ by the parameterization $\vec{P}_{k}(\lambda) = \big( x_{k}(\lambda), z_{k}(\lambda) \big) \ (\lambda_{\text{min}} \leq \lambda \leq \lambda_{\text{max}})$, then the rays that form the boundaries $\partial \text{S}^1_{j,k}$ and $\partial \text{T}^1_{k,j}$ are parameterized by

\begin{equation}\label{eq:2}
    \vec{r}^{1}_{j,k}(\lambda) = \vec{P}_{k}(\lambda) - \vec{P}_{j}(\gamma_{\text{min}}), \ (\lambda_{\text{min}} \leq \lambda \leq \lambda_{\text{max}}).
\end{equation}

The rays form a vertical segment in $\text{S}_{j}$ as only the direction coordinate changes; they form a curved segment in $\text{T}_{k}$ because both the position and direction coordinates change. The analytic expressions for $\partial \text{S}^1_{j,k}$ and $\partial \text{T}^1_{k,j}$ are

\begin{subequations}\label{eq:3}
    \begin{align}
        \partial \text{S}^{1}_{j,k}(\lambda) &= \Big\{ \big(x_{j}(\gamma_{\text{min}}), \ \unitvec{\tau}_{j}(\gamma_{\text{min}}) \dot \unitvec{r}^{1}_{j,k}(\lambda) \big) \ \Big| \ \lambda_{\text{min}} \leq \lambda \leq \lambda_{\text{max}} \Big\},\\
        \partial \text{T}^{1}_{k,j}(\lambda) &= \Big\{ \big(x_{k}(\lambda), \  -\unitvec{\tau}_k(\lambda) \dot \unitvec{r}^{1}_{j,k}(\lambda) \big) \ \Big| \ \lambda_{\text{min}} \leq \lambda \leq \lambda_{\text{max}} \Big\}.
    \end{align}
\end{subequations}

We indicate with $\unitvec{r}^{1}_{j,k}(\lambda)$ the normalization of the ray in Eq. (\ref{eq:2}) and with $\unitvec{\tau}_{j}(\gamma)$ and $\unitvec{\tau}_k(\lambda)$ the normalized tangent vectors to surfaces $j$ and $k$ respectively. The tangent vectors are obtained by rotating the inward facing surface normals $\unitvec{\nu}_j$ and $\unitvec{\nu}_k$ by an angle of $\pi/2$ counterclockwise. Note that the parameter in Eq. (\ref{eq:3}) corresponds to the surface that is traced out. The expressions for the other boundary segments are similar.

\subsection{Phase Space of the Two-Faceted Cup}\label{sec:2:1}
The two-faceted cup depicted in Fig. \ref{fig:cup} is a simple optical system consisting of four surfaces. A Lambertian light source (surface 1), two reflectors formed by straight line segments (surfaces 2, 3) and a target (surface 4). A ray leaving the source of the cup can reflect many times between the reflectors before reaching the target. Light that reflects on one of the reflectors always propagates to another surface. The phase spaces of the two-faceted cup can be seen in Fig. \ref{fig:phase_spaces_cup} and are given by the following expressions:

\vspace{-3mm}
\begin{equation}\label{eq:5}
    \begin{split}
        \text{S}_1 &= \text{S}_{1,2} \cup \text{S}_{1,3} \cup \text{S}_{1,4}, \quad \text{S}_2 = \text{S}_{2,3} \cup \text{S}_{2,4}, \quad \text{S}_3 = \text{S}_{3,2} \cup \text{S}_{3,4},\\
        \text{T}_2 &= \text{T}_{2,1} \cup \text{T}_{2,3}, \quad \text{T}_3 = \text{T}_{3,1} \cup \text{T}_{3,2}, \quad \text{T}_4 = \text{T}_{4,1} \cup \text{T}_{4,2} \cup \text{T}_{4,3}.
    \end{split}
\end{equation}

\subsection{Phase Space of the Compound Parabolic Concentrator}\label{sec:2:2}
The CPC \cite{chaves2008} depicted in Fig. \ref{fig:cpc} is a standard optical system that collects light from a Lambertian source and reshapes it to a focused beam. The system consists of an aperture (surface 4), a receiver (surface 1) and two reflectors that are segments of parabolas (surfaces 2, 3). We take the receiver as the light source and the aperture as the target, thus considering light traveling in the opposite direction. A light ray reflects on at most one reflector of the CPC. It can however reflect infinitely many times on a reflector. The parameterization of the right reflector follows from the polar equation of a parabola \cite{chaves2008} and is given by:

\begin{equation}\label{eq:6}
    \vec{P}(\phi) = \big( -h(\phi) \sin{(\phi + \theta)} - a, \ h(\phi) \cos{(\phi + \theta)} \big), \ \frac{3\pi}{2}-\theta \leq \phi \leq 2\pi-2\theta,
\end{equation}
where the axis of the parabola is rotated $\theta$ radians around the origin and the parabola is translated horizontally along a distance $a>0$. Note that the parabola has focus $(-a, 0)$ and intersects the horizontal axis at the point $(a,0)$. The parameterization of the left reflector is similar but it is rotated $-\theta$ radians, it is translated in the opposite direction and it has different bounds for $\phi$. The value of $h(\phi)$ is given by:

\begin{equation}\label{eq:7}
    h(\phi) = \frac{2a \ (1+\sin{(\theta)})}{1-\cos{(\phi)}}.
\end{equation}

\begin{figure}[H]
    \begin{subfigure}[t]{0.45\textwidth}
        \centering
        \includegraphics[width=\textwidth]{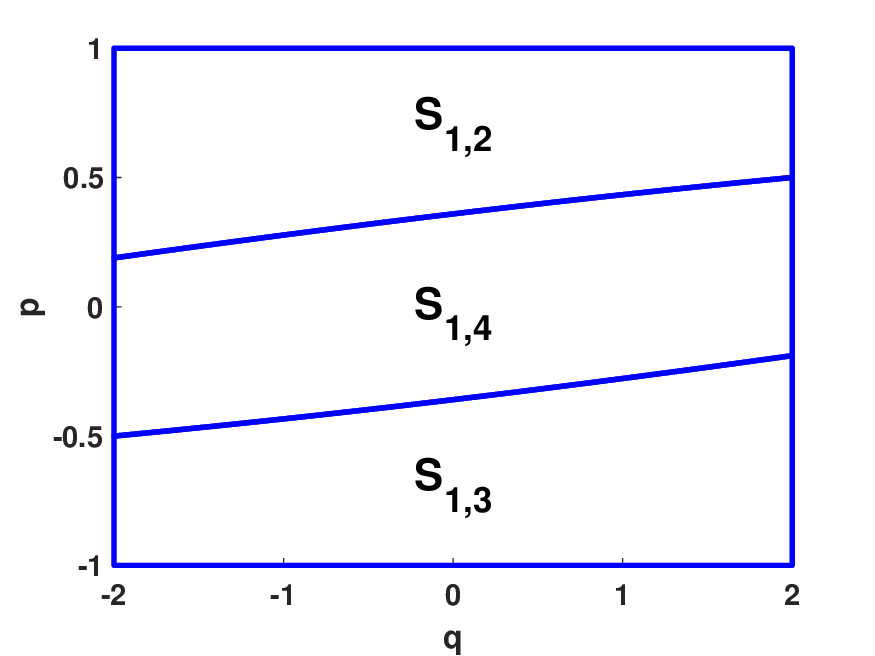}
        \caption{Source PS of surface 1 (light source). It is partitioned into regions $\text{S}_{1,k}$ where $k \in \{2,3,4\}$ containing rays emitted from surface 1 and reaching surface $k$.}
    \end{subfigure}
    \hspace{0.1\textwidth}
    \begin{subfigure}[t]{0.45\textwidth}
        \centering
        \includegraphics[width=\textwidth]{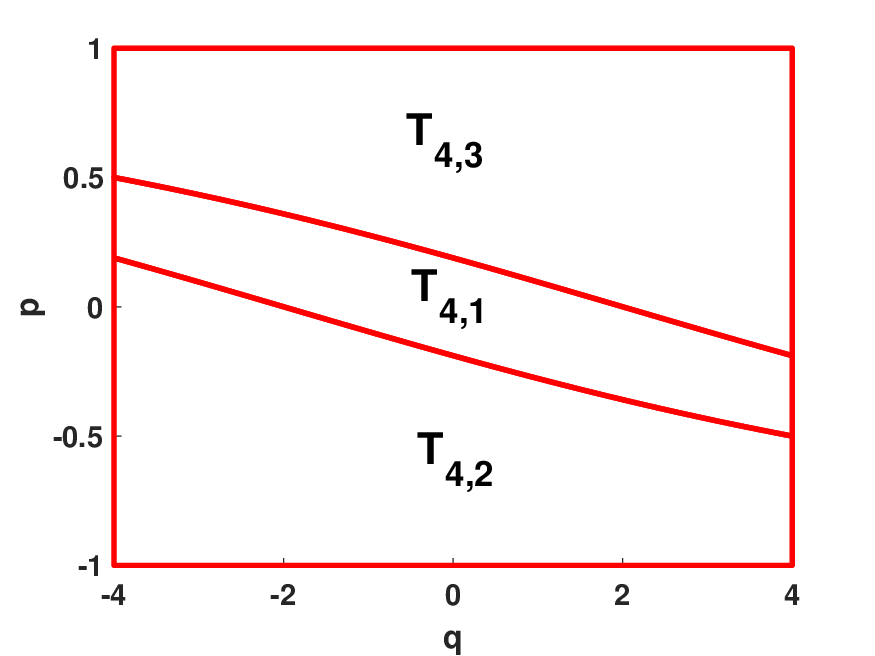}
        \caption{Target PS of surface 4 (target). It is partitioned into regions $\text{T}_{4,j}$ where $j \in \{1,2,3\}$ containing rays reaching surface 4 and emitted by surface $j$.}
    \end{subfigure}
    \begin{subfigure}[t]{0.45\textwidth}
        \centering
        \includegraphics[width=\textwidth]{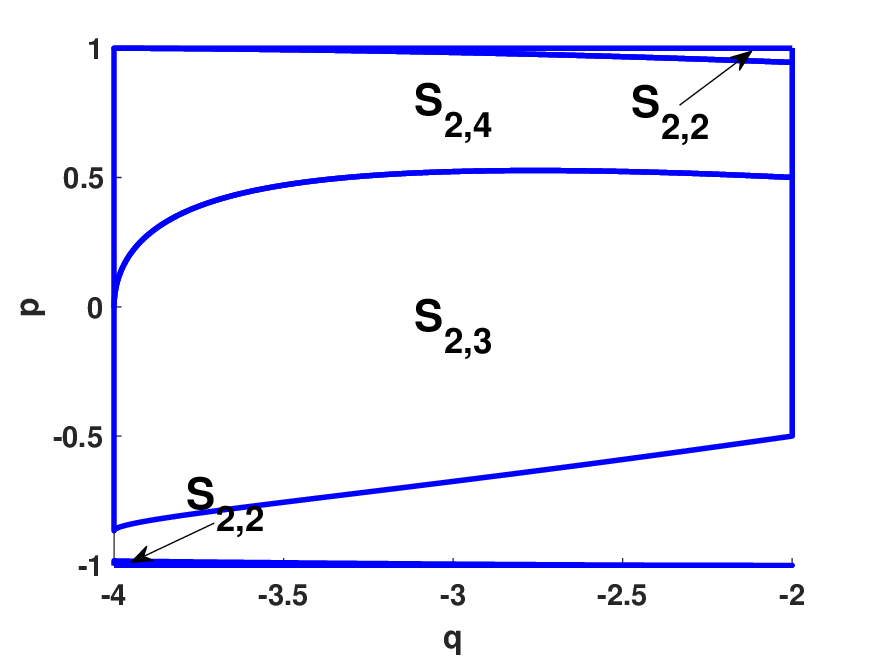}
        \caption{Source PS of surface 2 (left reflector). It is partitioned into regions $\text{S}_{2,k}$ where $k \in \{2,3,4\}$ containing rays emitted from surface 2 and reaching surface $k$.}
    \end{subfigure}
    \hspace{0.1\textwidth}
    \begin{subfigure}[t]{0.45\textwidth}
        \centering
        \includegraphics[width=\textwidth]{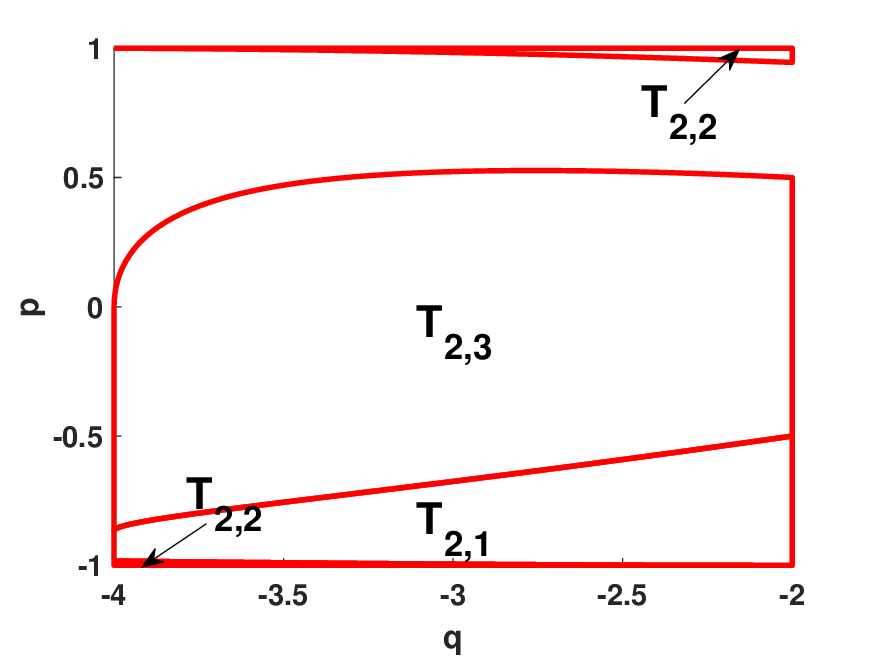}
        \caption{Target PS of surface 2 (left reflector). It is partitioned into regions $\text{T}_{2,j}$ where $j \in \{1,2,3\}$ containing rays reaching surface 2 and emitted by surface $j$.}
    \end{subfigure}
    \begin{subfigure}[t]{0.45\textwidth}
        \centering
        \includegraphics[width=\textwidth]{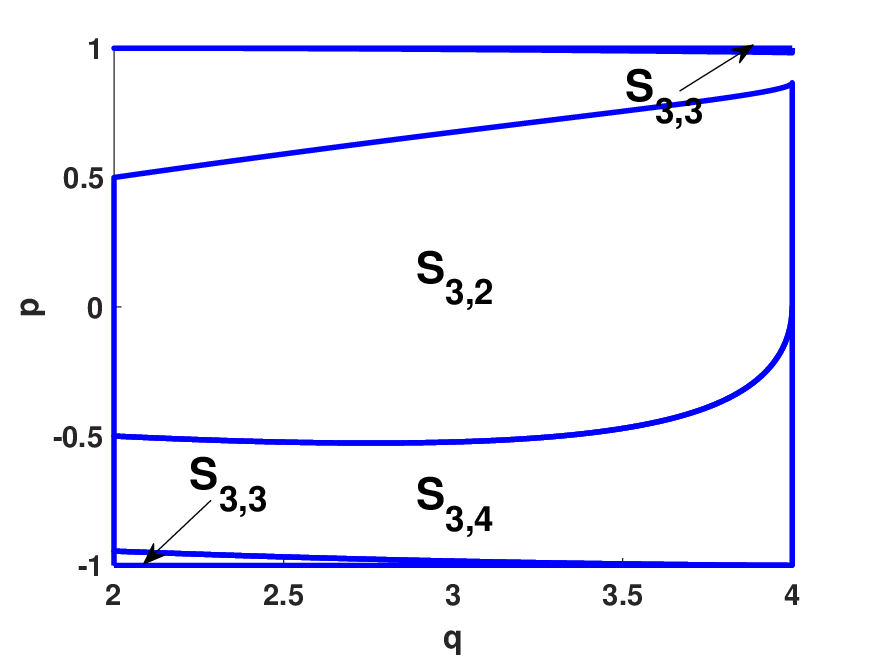}
        \caption{Source PS of surface 3 (right reflector). It is partitioned into regions $\text{S}_{3,k}$ where $k \in \{2,3,4\}$ containing rays emitted from surface 3 and reaching surface $k$.}
    \end{subfigure}
    \hspace{0.1\textwidth}
    \begin{subfigure}[t]{0.45\textwidth}
        \centering
        \includegraphics[width=\textwidth]{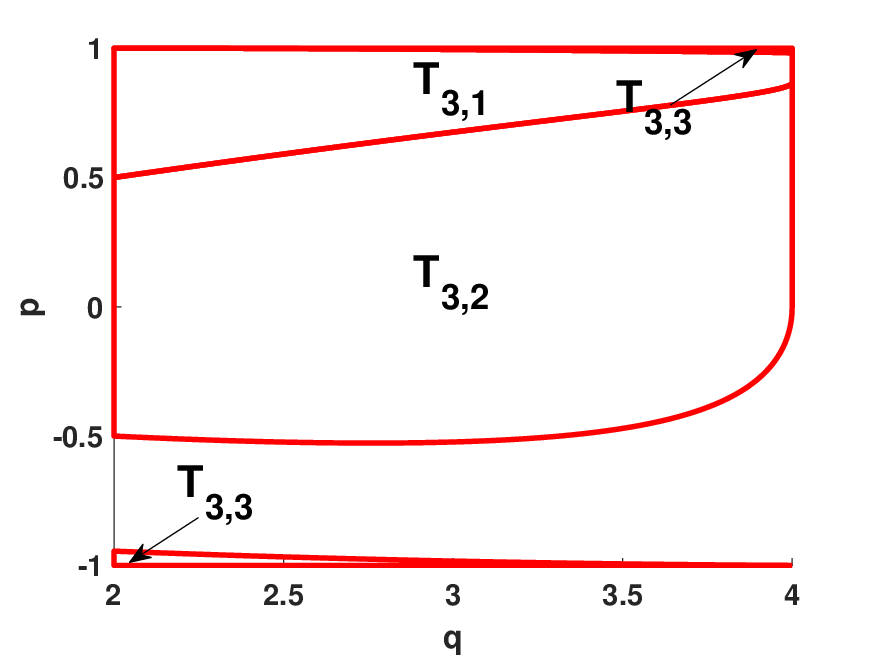}
        \caption{Target PS of surface 3 (right reflector). It is partitioned into regions $\text{T}_{3,j}$ where $j \in \{1,2,3\}$ containing rays reaching surface 3 and emitted by surface $j$.}
    \end{subfigure}
    \caption{Source and target phase spaces of the CPC.}
    \label{fig:phase_spaces_cpc}
\end{figure}

A ray leaving the source of the CPC can reflect many times on a single reflector before reaching the target. The phase spaces of the CPC can be seen in Fig. \ref{fig:phase_spaces_cpc} and are given by the following expressions:

\begin{equation}\label{eq:8}
    \begin{split}
        \text{S}_1 &= \text{S}_{1,2} \cup \text{S}_{1,3} \cup \text{S}_{1,4}, \quad \text{S}_2 = \text{S}_{2,2} \cup \text{S}_{2,3} \cup \text{S}_{2,4}, \quad \text{S}_3 = \text{S}_{3,2} \cup \text{S}_{3,3} \cup \text{S}_{3,4},\\
        \text{T}_2 &= \text{T}_{2,1} \cup \text{T}_{2,2} \cup \text{T}_{2,3}, \quad \text{T}_3 = \text{T}_{3,1} \cup \text{T}_{3,2} \cup \text{T}_{3,3}, \quad \text{T}_4 = \text{T}_{4,1} \cup \text{T}_{4,2} \cup \text{T}_{4,3}.
    \end{split}
\end{equation}

\section{Concatenated Backward Ray Mapping}\label{sec:3}
For each optical system there exists an optical map $\text{\textbf{M}}_{1,\text{N}} : \text{S}_{1} \to \text{T}_{\text{N}}$ such that $\text{\textbf{M}}_{1,\text{N}}(q_{\text{s}, 1}, p_{\text{s}, 1})=(q_{\text{t}, \text{N}}, p_{\text{t}, \text{N}})$ for every $(q_{\text{s}, 1}, p_{\text{s}, 1}) \in \text{S}_{1}$. All rays that follow the same path $\Pi$ from source to target form a region $\text{R}_{\text{s}}(\Pi) \subset \text{S}_1$ and $\text{R}_{\text{t}}(\Pi) \subset \text{T}_n$. A path is the sequence of surfaces encountered by a ray traveling from source to target. The map $\text{\textbf{M}}_{1,\text{N}}(\Pi)$ is the map $\text{\textbf{M}}_{1,\text{N}}$ restricted to the path $\Pi$ and relates $\text{R}_{\text{s}}(\Pi)$ to $\text{R}_{\text{t}}(\Pi)$. The areas covered by light rays in $\text{S}_1$ and $\text{T}_n$ are equal because of étendue conservation \cite{chaves2008}. Empty regions occur where no light rays exist that travel from source to target.

The light intensity $I(p)$ at the target for a given $p=\text{const}$ is computed by integrating the target luminance over the position coordinate $q$ and is defined by:

\begin{equation}\label{eq:9}
    I(p) = \int_{Q} L(q, p) \, \text{d}q.
\end{equation}
The light intensity in $\text{T}_n$ depends on the luminance, which is positive in non-empty regions of PS. Assuming positive luminance on the source gives the following relation:

\begin{figure}[H]
    \vspace{-6mm}
    \begin{subfigure}[t]{0.45\textwidth}
        \centering
        \includegraphics[width=\textwidth]{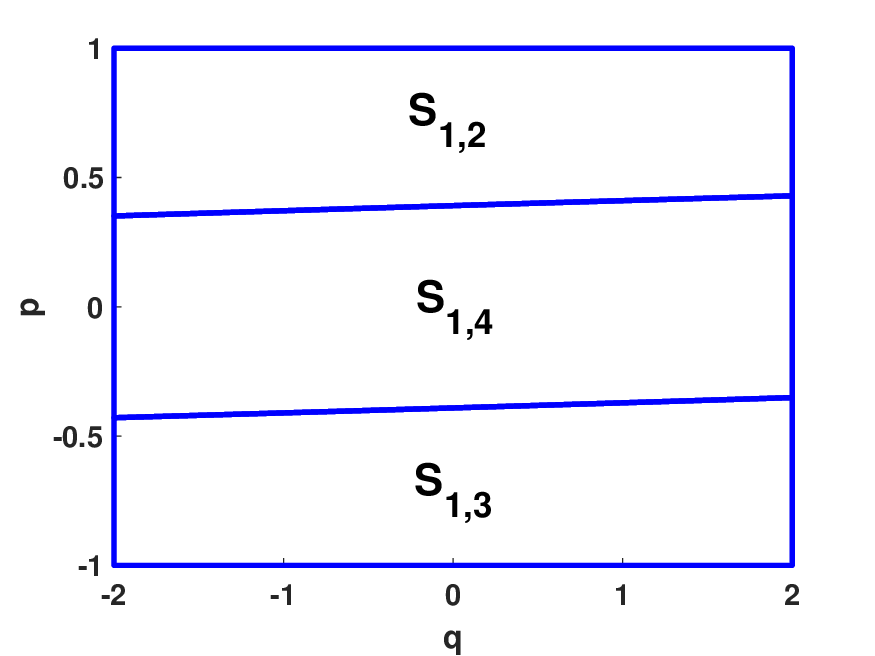}
        \caption{Source PS of surface 1 (light source). It is partitioned into regions $\text{S}_{1,k}$ where $k \in \{2,3,4\}$ containing rays emitted from surface 1 and reaching surface $k$.}
    \end{subfigure}
    \hspace{0.1\textwidth}
    \begin{subfigure}[t]{0.45\textwidth}
        \centering
        \includegraphics[width=\textwidth]{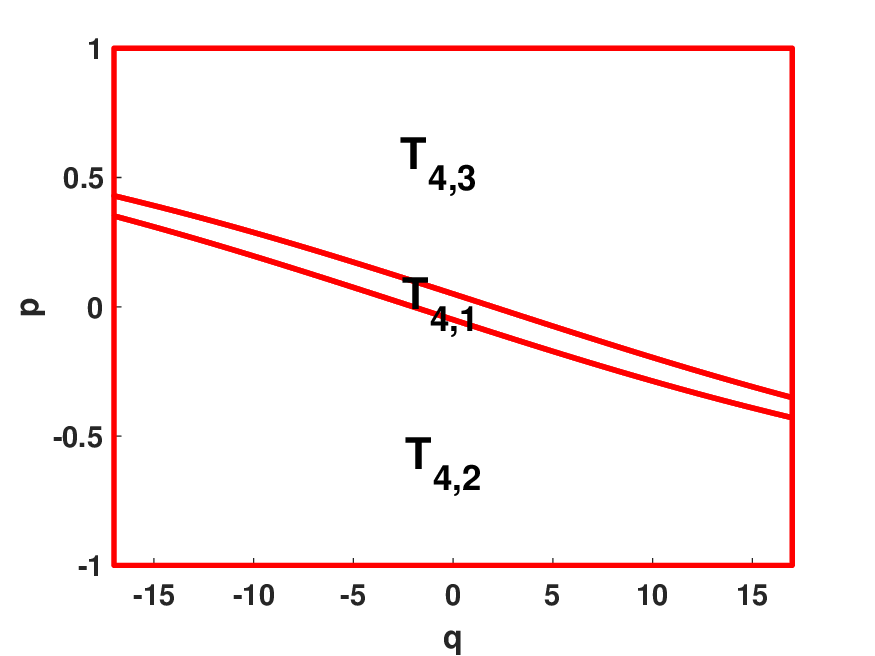}
        \caption{Target PS of surface 4 (target). It is partitioned into regions $\text{T}_{4,j}$ where $j \in \{1,2,3\}$ containing rays reaching surface 4 and emitted by surface $j$.}
    \end{subfigure}
    \begin{subfigure}[t]{0.45\textwidth}
        \centering
        \includegraphics[width=\textwidth]{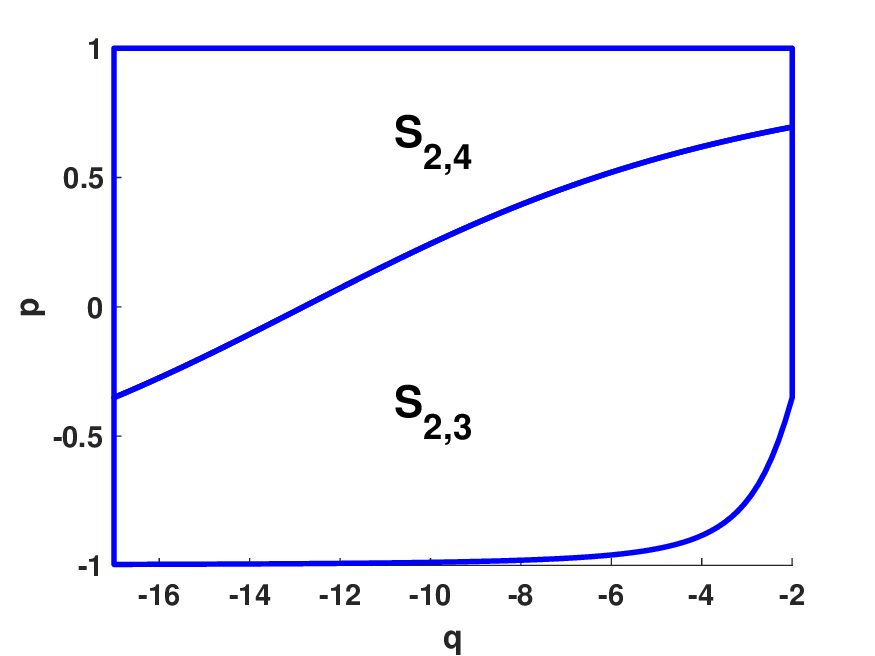}
        \caption{Source PS of surface 2 (left reflector). It is partitioned into regions $\text{S}_{2,k}$ where $k \in \{3,4\}$ containing rays emitted from surface 2 and reaching surface $k$.}
    \end{subfigure}
    \hspace{0.1\textwidth}
    \begin{subfigure}[t]{0.45\textwidth}
        \centering
        \includegraphics[width=\textwidth]{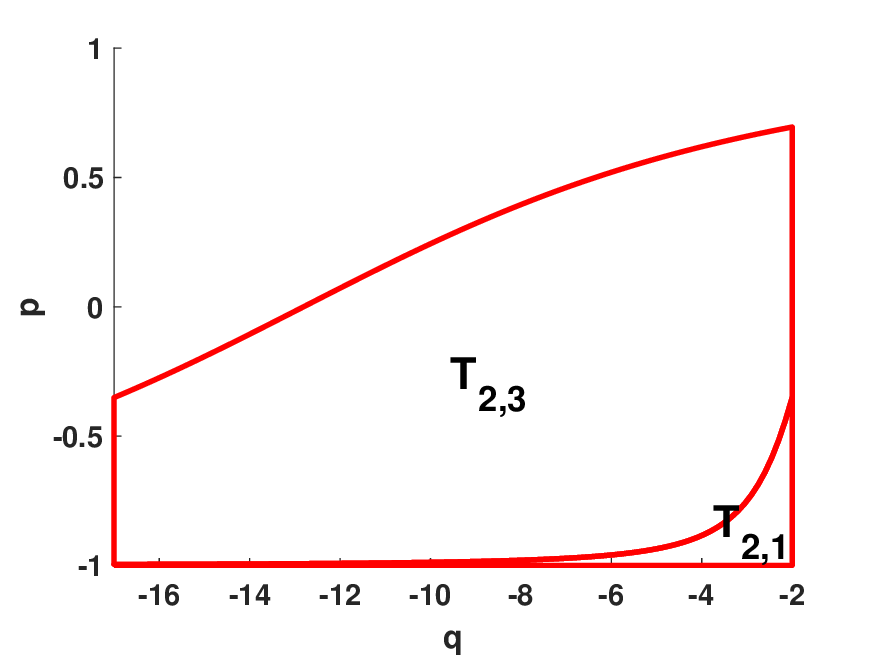}
        \caption{Target PS of surface 2 (left reflector). It is partitioned into regions $\text{T}_{2,j}$ where $j \in \{1,3\}$ containing rays reaching surface 2 and emitted by surface $j$.}
    \end{subfigure}
    \begin{subfigure}[t]{0.45\textwidth}
        \centering
        \includegraphics[width=\textwidth]{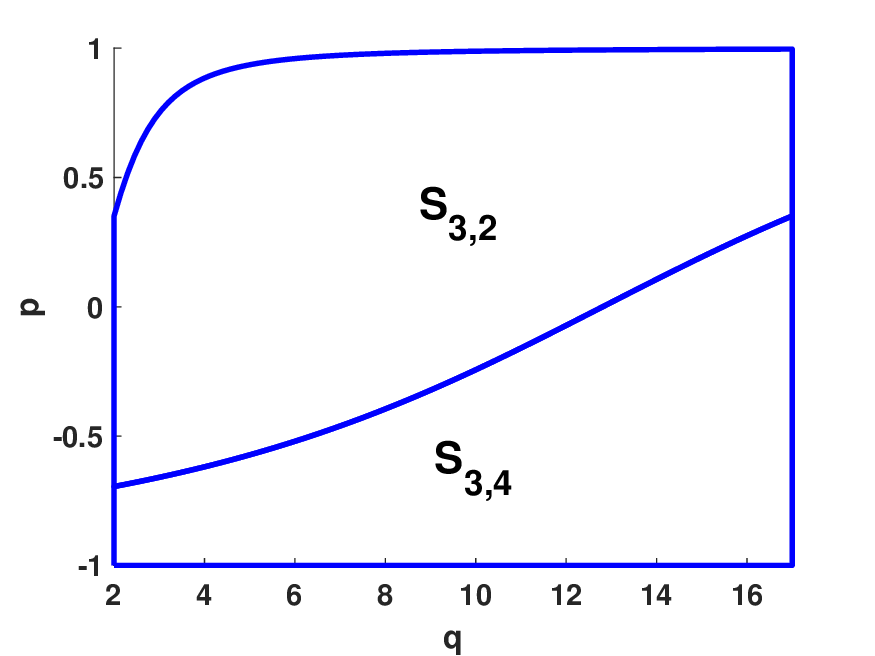}
        \caption{Source PS of surface 3 (right reflector). It is partitioned into regions $\text{S}_{3,k}$ where $k \in \{2,4\}$ containing rays emitted from surface 3 and reaching surface $k$.}
    \end{subfigure}
    \hspace{0.1\textwidth}
    \begin{subfigure}[t]{0.45\textwidth}
        \centering
        \includegraphics[width=\textwidth]{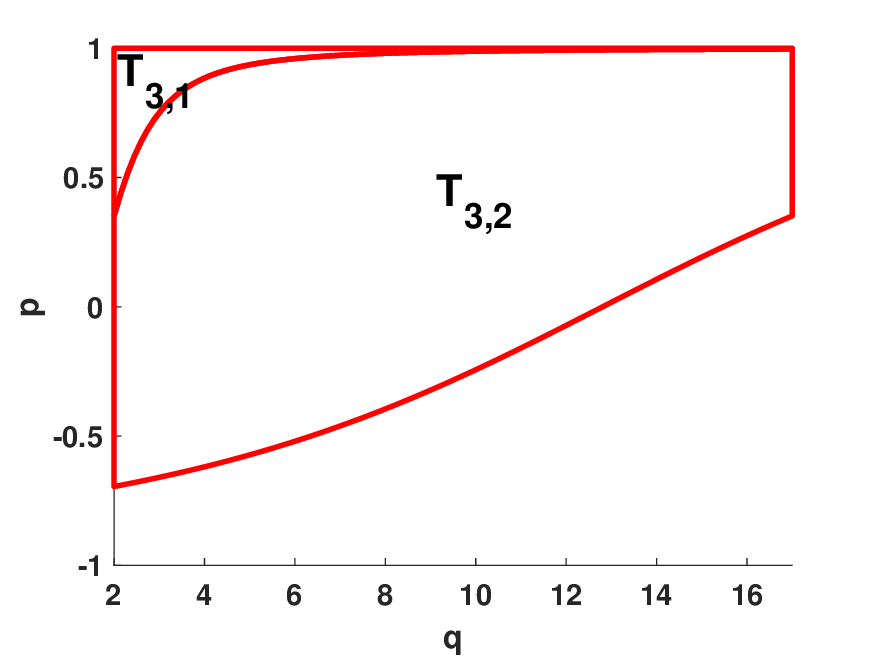}
        \caption{Target PS of surface 3 (right reflector). It is partitioned into regions $\text{T}_{3,j}$ where $j \in \{1,2\}$ containing rays reaching surface 3 and emitted by surface $j$.}
    \end{subfigure}
    \caption{Source and target phase spaces of the two-faceted cup.}
    \label{fig:phase_spaces_cup}
\end{figure}

\begin{equation}\label{eq:10}
    \begin{split}
        L(q,p) &> 0 \qquad \forall_{q,p} \in \text{T}_{\text{N},1},\\
        L(q,p) &\geq 0 \qquad \forall_{q,p} \in \text{T}_{\text{N},j}, \ j \in \{2,\dots,\text{N}-1\}.
    \end{split}
\end{equation}

The phase spaces of an optical system are connected through maps that relate the coordinates on every PS. Propagation maps $\text{\textbf{P}}_{j,k} : \text{S}_{j,k} \to \text{T}_{k,j}$ describe light that travels from a surface $j$ to another surface $k$; they relate coordinates of $\text{S}_j$ to $\text{T}_k$ such that $\text{\textbf{P}}_{j,k}(q_{\text{s}, j}, p_{\text{s}, j}) = (q_{\text{t}, k}, p_{\text{t}, k})$. Reflection maps $\text{\textbf{R}}_{k} : \text{T}_{k} \to \text{S}_{k}$ describe light that reflects on a surface $k$; they relate coordinates of $\text{T}_k$ to $\text{S}_k$ such that $\text{\textbf{R}}_{k}(q_{\text{t}, k}, p_{\text{t}, k}) = (q_{\text{s}, k}, p_{\text{s}, k})$. Note that $q_{\text{t}, k} = q_{\text{s}, k}$. Every map $\text{\textbf{M}}_{1,\text{N}}(\Pi)$ can be described by a composition of propagation and reflection maps. Considering all paths $\Pi$ from source to target, the positive luminance regions $\text{R}_{\text{t}}(\Pi) \subset \text{T}_n$ can be determined. From Eq. (\ref{eq:10}) follows that the luminance for some path $\Pi$ connecting the source and target is:

\begin{equation}\label{eq:11}
    \begin{split}
        L(q,p) &> 0 \qquad \forall_{q,p} \in \text{R}_{\text{t}}(\Pi),\\
        L(q,p) &=0 \qquad \text{otherwise}.
    \end{split}
\end{equation}

A Lambertian source emits light with a constant luminance \cite{chaves2008}. Assuming a Lambertian source with luminance equal to 1, computing $I(p)$ reduces to computing the boundaries $\partial\text{R}_{\text{t}}(\Pi)$ of the regions of all possible paths $\Pi$. The intensity is given by the sum of the interval lengths formed by the intersections of the support of the luminance and the line $p=\text{const}$. The intersection points between $p=\text{const}$ and the boundary $\partial\text{R}_{\text{t}}(\Pi)$ have position coordinates $q^{\text{min}}(\Pi, p)$ and $q^{\text{max}}(\Pi, p)$, where $q^{\text{min}}(\Pi, p) < q^{\text{max}}(\Pi, p)$. Using Eq. (\ref{eq:11}), it follows that Eq. (\ref{eq:9}) reduces to:

\begin{equation}\label{eq:12}
    I(p) = \sum_\Pi \int^{q^{\text{max}}(\Pi, p)}_{q^{\text{min}}(\Pi, p)} L(q, p) \, \text{d$q$} = \sum_\Pi(q^{\text{max}}(\Pi, p) - q^{\text{min}}(\Pi, p)).
\end{equation}

CBRM computes the light intensity $I(p)$ using the phase spaces of all surfaces of an optical system. A parallel light beam is represented by a straight line segment on the line $p=\text{const}$ in the source/target PS of a straight surface since the direction coordinate of all rays is the same. The intersections between $p=\text{const}$ and the boundaries in PS are computed several times by the algorithm; this is done analytically since $p=\text{const}$ is a horizontal line, and we have analytical descriptions of all boundaries of all phase spaces. Therefore, the light beam is required to always stay parallel which in turn requires the optical system to consist of only straight surfaces.

The algorithm uses the map $\text{\textbf{M}}_{1,\text{N}}(\Pi) : \text{R}_{\text{s}}(\Pi) \to \text{R}_{\text{t}}(\Pi)$ for all possible paths $\Pi$ to compute $I(p)$ for a given direction $p=\text{const}$ in $\text{T}_n$. To construct the map $\text{\textbf{M}}_{1,\text{N}}(\Pi)$, the corresponding path $\Pi$ should be known. The algorithm computes all possible paths $\Pi$ by considering rays in $\text{T}_n$ along a given direction $p \in [-1,1]$ and tracing them backward recursively in PS \cite{Filosa2021}. The endpoints of the light beam in $\text{T}_n$ are $(q^{\text{min}}_{\text{t},\text{N}}, p_{\text{t},\text{N}})$ and $(q^{\text{max}}_{\text{t},\text{N}}, p_{\text{t},\text{N}})$. The line $p=\text{const}$ intersects various regions in PS. The intersection points with boundaries $\partial \text{T}_{\text{N},j}$ are $(u^{\text{min}}_{\text{N},j}, p_{\text{t},\text{N}})$ and $(u^{\text{max}}_{\text{N},j}, p_{\text{t},\text{N}})$. The intersection segment with region $\text{T}_{\text{N},j}$ is given by $[v^{\text{min}}_{\text{N},j}, v^{\text{max}}_{\text{N},j}] = [q^{\text{min}}_{\text{t},\text{N}}, q^{\text{max}}_{\text{t},\text{N}}] \cap [u^{\text{min}}_{\text{N},j}, u^{\text{max}}_{\text{N},j}]$ and corresponds to rays emitted by another surface $j \neq \text{N}$. The endpoints of the intervals are transformed to coordinates $(q^{\text{min}}_{\text{t},j}, p_{\text{t},j})$ and $(q^{\text{max}}_{\text{t},j}, p_{\text{t},j})$ in $\text{T}_j$ by sequentially applying the maps $\text{\textbf{P}}^{-1}_{j,\text{N}} : \text{T}_{\text{N},j} \to \text{S}_{j,\text{N}}$ and $\text{\textbf{R}}^{-1}_{j} : \text{S}_{j} \to \text{T}_{j}$. The procedure is repeated in $\text{T}_j$. The recursion ends when an intersection segment $[v^{\text{min}}_{k,j}, v^{\text{max}}_{k,j}]$ is traced back to $\text{S}_1$ or when an intersection segment is empty. If $\text{S}_1$ is found, the endpoints $(q^{\text{min}}_{\text{s},1}, p_{\text{s},1})$ and $(q^{\text{max}}_{\text{s},1}, p_{\text{s},1})$ of the light beam are traced to $\text{T}_n$ along $\Pi$ by applying $\text{\textbf{M}}_{1,\text{N}}(\Pi)$. This gives two points $(q^{\text{min}}(\Pi, p), p)$ and $(q^{\text{max}}(\Pi, p), p)$ at the boundary of a region $\text{R}_{\text{t}}(\Pi) \subset \text{T}_n$ with positive luminance. The main steps to calculate $I(p)$ are given in Algorithm \ref{alg:cbrm}.

The range of angular coordinates at the target is divided into $\text{Ni}$ equidistant intervals with endpoints $p^{m}$ and $p^{m+1}$ where $m \in \{0, \dots, \text{Ni}-1 \}$. The averaged and normalized intensity $\hat{I}$ is given for every interval $p^{m+1/2} = \frac{1}{2}(p^{m}+p^{m+1})$ by:

\begin{equation}\label{eq:13}
    \hat{I}(p^{m + 1/2}) = \frac{1}{U_{\text{t}}} \int_{p_{m}}^{p_{m+1}} I(p) \, \text{d}p,
\end{equation}
and is computed using the trapezoidal rule. $U_{\text{t}}$ denotes the total étendue at the target, which in PS corresponds to the area of the region covered by the light rays \cite{FilosaThesis}. The intensity distribution is obtained by plotting $p^{m+1/2}$ on the horizontal axis and $\hat{I}(p^{m + 1/2})$ on the vertical axis for each interval. For more details on the CBRM algorithm see \cite{Filosa2021, FilosaThesis}.

Fig. \ref{fig:algo} shows the first steps of the algorithm for light on $p=-0.2$ in $\text{T}_4$ on the two-faceted cup. In step 1 (Fig. \ref{fig:algo_1}) we find light traveling directly from source to target. The algorithm updates the intensity and continues to the next region of PS in $\text{T}_4$. In step 2 (Fig. \ref{fig:algo_2}) we find light traveling from surface 2 (left reflector) to the target. The light is traced to $\text{T}_2$ where it lies on $\text{p}=-0.82$ and the procedure is repeated. In step 3 (Fig. \ref{fig:algo_3}) we find light traveling from the source to surface 2. The algorithm traces this light back to the target and updates the intensity, then it continues to the next region of $\text{T}_2$. In step 4 (Fig. \ref{fig:algo_4}) we find light traveling from surface 3 (right reflector) to surface 2. The light is traced to $\text{T}_3$ where it lies on $\text{p}=0.29$ and the procedure is repeated. In step 5 (Fig. \ref{fig:algo_5}) we find light traveling from surface 2 to surface 3. The light is traced to $\text{T}_2$ where it lies on $\text{p}=0.41$ and the procedure is repeated. In step 6 (Fig. \ref{fig:algo_6}) we find light traveling from surface 3 to surface 2. The light is traced to $\text{T}_3$ where it does not intersect any region of PS meaning it was not emitted by the source. Therefore, the computation for this part of the light beam stops; the algorithm is finished for the light of step 2 reaching the target from surface 2. The process is repeated for the light reaching the target from surface 3.

\subsection{Intensity Distribution of the Two-Faceted Cup}\label{sec:3:1}
\begin{figure}[htb]
    \begin{minipage}{0.45\textwidth}
        \centering
        \includegraphics[width=\textwidth]{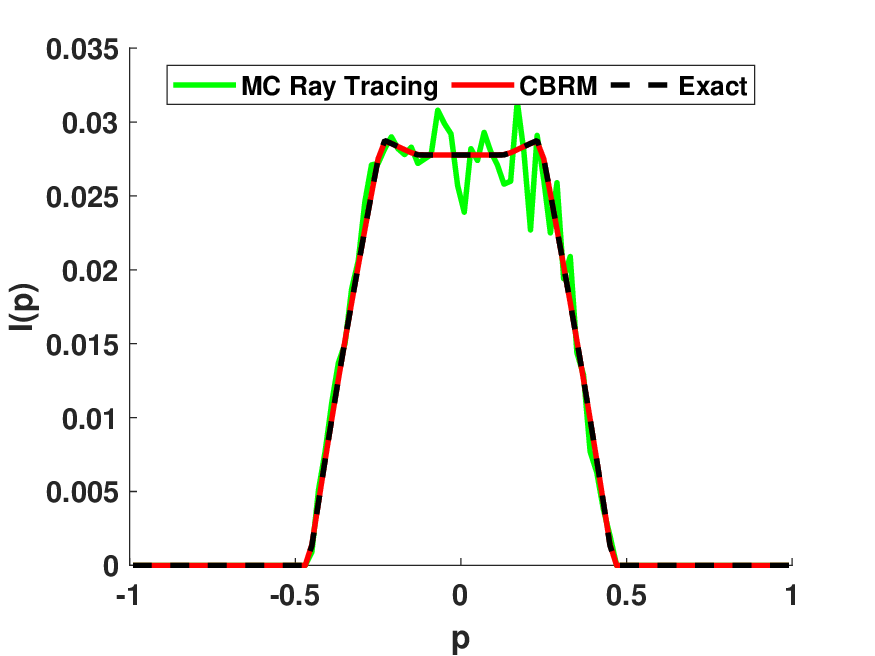}
        \caption{Intensity distributions of the two-faceted cup computed with MC ray tracing (green) and CBRM (red) compared to the exact solution.}
        \label{fig:intensity_cup}
    \end{minipage}
    \hspace{0.1\textwidth}
    \begin{minipage}{0.45\textwidth}
        \centering
        \includegraphics[width=\textwidth]{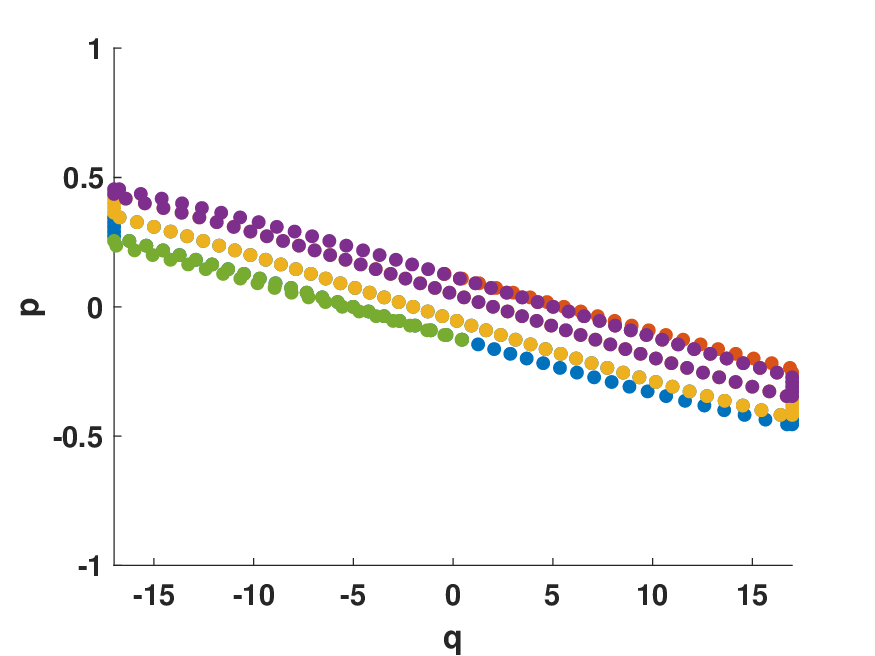}
        \caption{Points on the boundaries of the positive luminance regions in $\text{T}_4$ of the cup for all paths $\Pi$. Each color represents a different path.}
        \label{fig:backward_ps_cup}
    \end{minipage}
\end{figure}

The cup is a simple system for which we can compute the intensity distribution exactly \cite{swi_2012}. The target is divided into 100 bins for (Q)MC ray tracing. The intensity in each bin is also computed with CBRM using Eq. (\ref{eq:13}). The intensity distribution found with MC ray tracing, shown in Fig. \ref{fig:intensity_cup}, is noisy and not close to the exact solution. The intensity distribution computed with CBRM, shown in Fig. \ref{fig:intensity_cup}, matches the exact solution precisely. MC ray tracing required $10^4$ rays, but CBRM required only $10^3$ rays. CBRM is more accurate than MC ray tracing and requires fewer rays to compute the intensity. We can use CBRM to find the boundaries of the positive luminance regions in $\text{T}_4$; they are shown in Fig. \ref*{fig:backward_ps_cup}.

\begin{figure}[H]
    \begin{subfigure}[t]{0.45\textwidth}
        \centering
        \includegraphics[width=\textwidth]{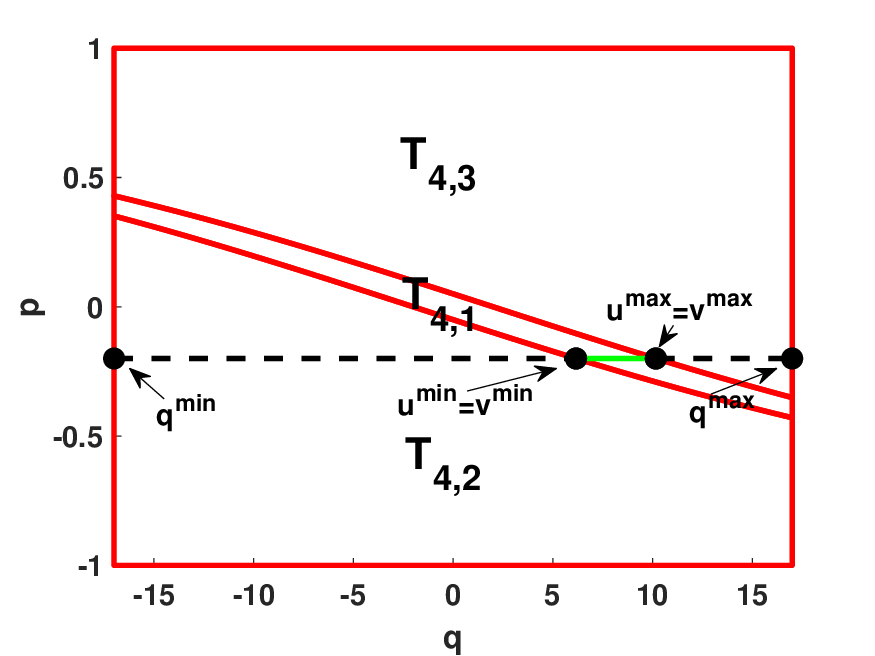}
        \caption{Light in $\text{T}_{4}$ on the line $p=-0.2$. $(q^{\text{min}}_{\text{t},4}, p)$ and $(q^{\text{max}}_{\text{t},4}, p)$ are the endpoints of the light beam. The intersection points between $p=-0.2$ and $\partial\text{T}_{4,1}$ are $(u^{\text{min}}_{4,1}, p)$ and $(u^{\text{max}}_{4,1}, p)$. The intersection segment has endpoints $(v^{\text{min}}_{4,1}, p)$ and $(v^{\text{max}}_{4,1}, p)$ with $v^{\text{min}}_{4,1}=\max\{q^{\text{min}}_{\text{t},4}, u^{\text{min}}_{4,1}\}$ and $v^{\text{max}}_{4,1}=\min\{q^{\text{max}}_{\text{t},4}, u^{\text{max}}_{4,1}\}$.}
        \label{fig:algo_1}
    \end{subfigure}
    \hspace{0.1\textwidth}
    \begin{subfigure}[t]{0.45\textwidth}
        \centering
        \includegraphics[width=\textwidth]{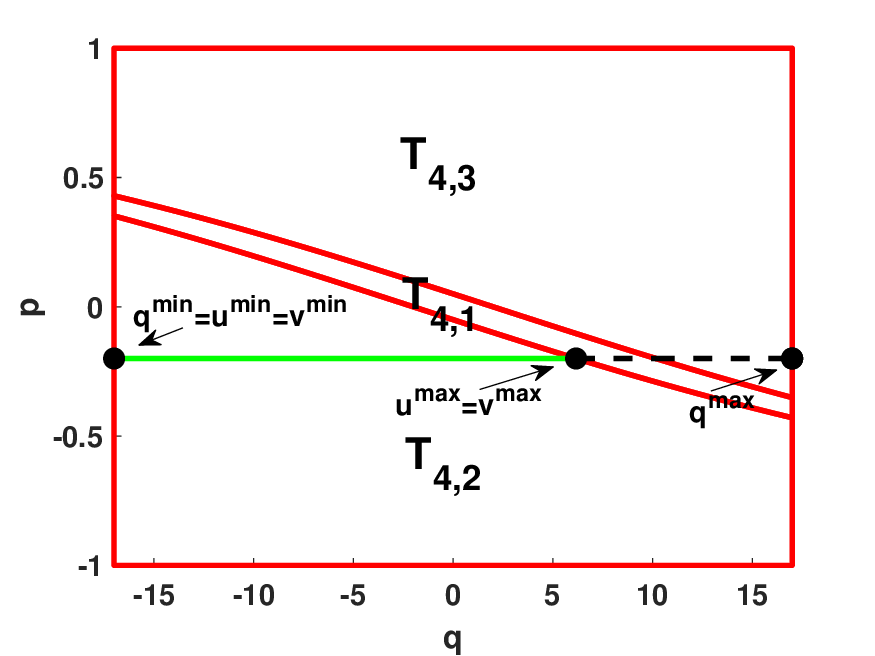}
        \caption{Light in $\text{T}_{4}$ on the line $p=-0.2$. $(q^{\text{min}}_{\text{t},4}, p)$ and $(q^{\text{max}}_{\text{t},4}, p)$ are the endpoints of the light beam. The intersection points between $p=-0.2$ and $\partial\text{T}_{4,2}$ are $(u^{\text{min}}_{4,2}, p)$ and $(u^{\text{max}}_{4,2}, p)$. The intersection segment has endpoints $(v^{\text{min}}_{4,2}, p)$ and $(v^{\text{max}}_{4,2}, p)$ with $v^{\text{min}}_{4,2}=\max\{q^{\text{min}}_{\text{t},4}, u^{\text{min}}_{4,2}\}$ and $v^{\text{max}}_{4,2}=\min\{q^{\text{max}}_{\text{t},4}, u^{\text{max}}_{4,2}\}$.}
        \label{fig:algo_2}
    \end{subfigure}
    \caption{Concatenated backward ray mapping on the two-faceted cup for $p=-0.2$ in $\text{T}_4$. The intersection segment computed at each step is colored green.}
\end{figure}
\begin{figure}[H]\ContinuedFloat
    \begin{subfigure}[t]{0.45\textwidth}
        \centering
        \includegraphics[width=\textwidth]{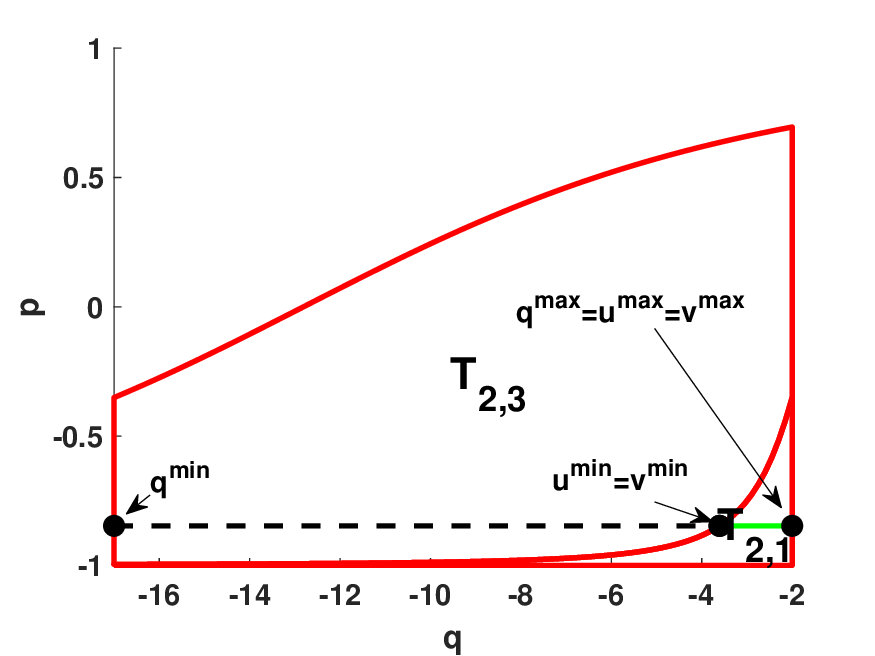}
        \caption{Light in $\text{T}_{2}$ on the line $p=-0.82$. $(q^{\text{min}}_{\text{t},2}, p)$ and $(q^{\text{max}}_{\text{t},2}, p)$ are the endpoints of the light beam. The intersection points between $p=-0.82$ and $\partial\text{T}_{2,1}$ are $(u^{\text{min}}_{2,1}, p)$ and $(u^{\text{max}}_{2,1}, p)$. The intersection segment has endpoints $(v^{\text{min}}_{2,1}, p)$ and $(v^{\text{max}}_{2,1}, p)$ with $v^{\text{min}}_{2,1}=\max\{q^{\text{min}}_{\text{t},2}, u^{\text{min}}_{2,1}\}$ and $v^{\text{max}}_{2,1}=\min\{q^{\text{max}}_{\text{t},2}, u^{\text{max}}_{2,1}\}$.}
        \label{fig:algo_3}
    \end{subfigure}
    \hspace{0.1\textwidth}
    \begin{subfigure}[t]{0.45\textwidth}
        \centering
        \includegraphics[width=\textwidth]{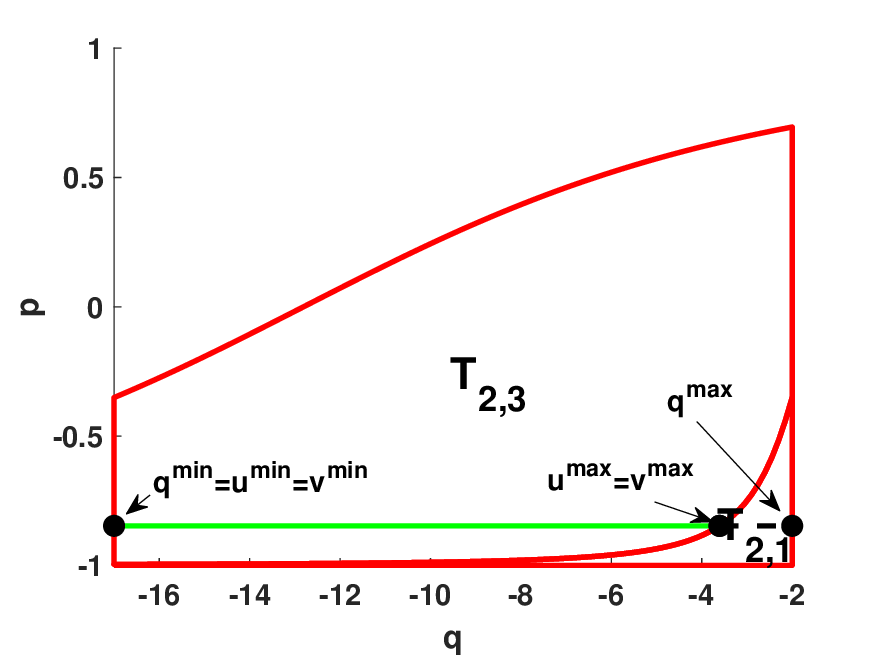}
        \caption{Light in $\text{T}_{2}$ on the line $p=-0.82$. $(q^{\text{min}}_{\text{t},2}, p)$ and $(q^{\text{max}}_{\text{t},2}, p)$ are the endpoints of the light beam. The intersection points between $p=-0.82$ and $\partial\text{T}_{2,3}$ are $(u^{\text{min}}_{2,3}, p)$ and $(u^{\text{max}}_{2,3}, p)$. The intersection segment has endpoints $(v^{\text{min}}_{2,3}, p)$ and $(v^{\text{max}}_{2,3}, p)$ with $v^{\text{min}}_{2,3}=\max\{q^{\text{min}}_{\text{t},2}, u^{\text{min}}_{2,3}\}$ and $v^{\text{max}}_{2,3}=\min\{q^{\text{max}}_{\text{t},2}, u^{\text{max}}_{2,3}\}$.}
        \label{fig:algo_4}
    \end{subfigure}
    \begin{subfigure}[t]{0.45\textwidth}
        \centering
        \includegraphics[width=\textwidth]{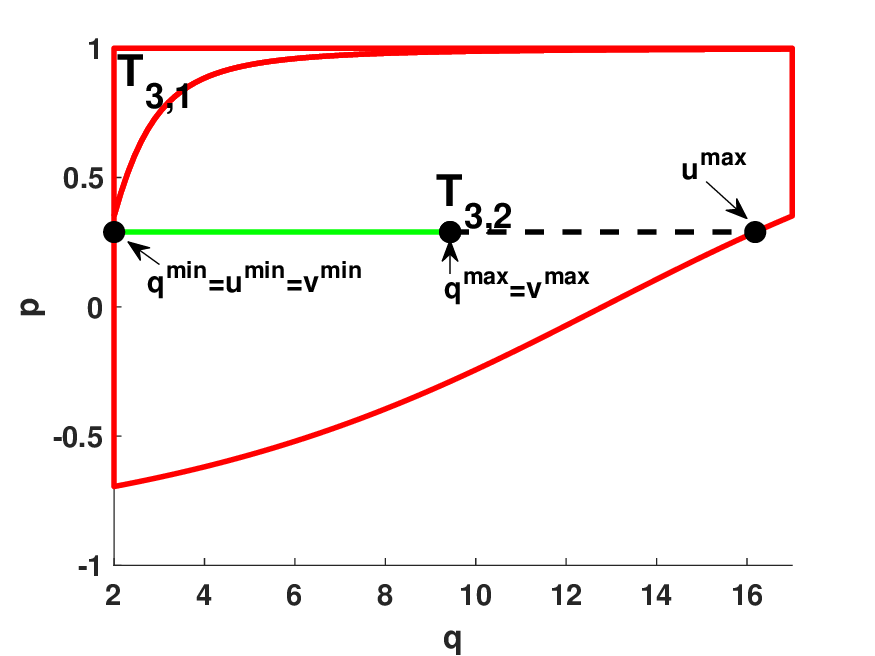}
        \caption{Light in $\text{T}_{3}$ on the line $p=0.29$. $(q^{\text{min}}_{\text{t},3}, p)$ and $(q^{\text{max}}_{\text{t},3}, p)$ are the endpoints of the light beam. The intersection points between $p=0.29$ and $\partial\text{T}_{3,2}$ are $(u^{\text{min}}_{3,2}, p)$ and $(u^{\text{max}}_{3,2}, p)$. The intersection segment has endpoints $(v^{\text{min}}_{3,2}, p)$ and $(v^{\text{max}}_{3,2}, p)$ with $v^{\text{min}}_{3,2}=\max\{q^{\text{min}}_{\text{t},3}, u^{\text{min}}_{3,2}\}$ and $v^{\text{max}}_{3,2}=\min\{q^{\text{max}}_{\text{t},3}, u^{\text{max}}_{3,2}\}$.}
        \label{fig:algo_5}
    \end{subfigure}
    \hspace{0.1\textwidth}
    \begin{subfigure}[t]{0.45\textwidth}
        \centering
        \includegraphics[width=\textwidth]{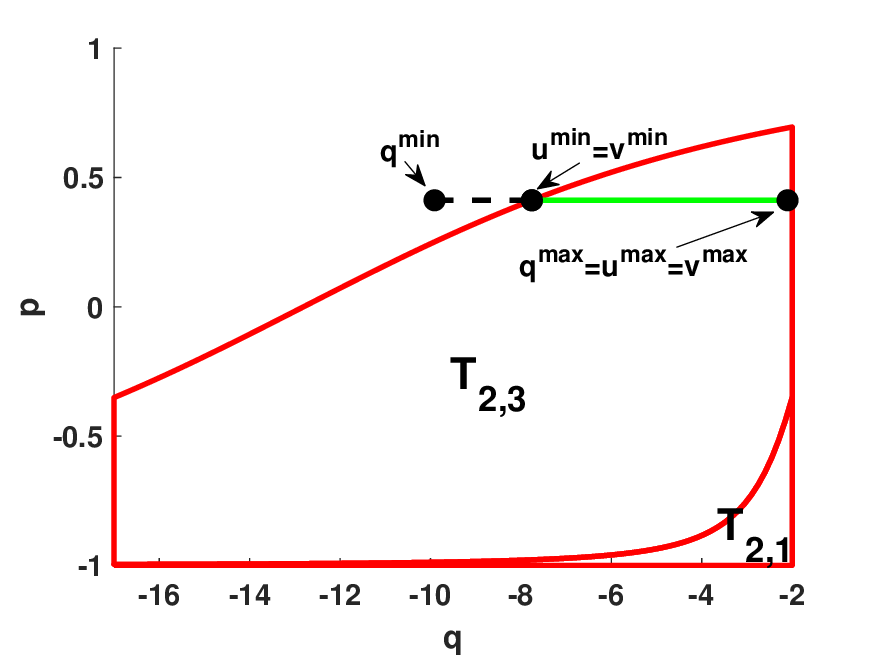}
        \caption{Light in $\text{T}_{2}$ on the line $p=0.41$. $(q^{\text{min}}_{\text{t},2}, p)$ and $(q^{\text{max}}_{\text{t},2}, p)$ are the endpoints of the light beam. The intersection points between $p=0.41$ and $\partial\text{T}_{2,3}$ are $(u^{\text{min}}_{2,3}, p)$ and $(u^{\text{max}}_{2,3}, p)$. The intersection segment has endpoints $(v^{\text{min}}_{2,3}, p)$ and $(v^{\text{max}}_{2,3}, p)$ with $v^{\text{min}}_{2,3}=\max\{q^{\text{min}}_{\text{t},2}, u^{\text{min}}_{2,3}\}$ and $v^{\text{max}}_{2,3}=\min\{q^{\text{max}}_{\text{t},2}, u^{\text{max}}_{2,3}\}$.}
        \label{fig:algo_6}
    \end{subfigure}
    \caption{Concatenated backward ray mapping on the two-faceted cup, continued.}
    \label{fig:algo}
\end{figure}

\begin{algorithm}[htb]
    \caption{Recursive procedure to compute the intensity for optical systems consisting of only straight surfaces}
    \textbf{Input}: $k=\text{N}, q^{\text{min}}_{\text{t},k}$ is the $x$-coordinate of the left endpoint of surface $\text{N}, q^{\text{max}}_{\text{t},k}$ is the $x$-coordinate of the right endpoint of surface $\text{N}, p_{\text{t},k}=p=\text{const}, I(p)=0, \Pi=(\text{N})$.
    \begin{algorithmic}[1]
    \Procedure{INTENSITY COMPUTATION}{$k, q^{\text{min}}_{\text{t},k}, q^{\text{max}}_{\text{t},k}, p, I(p), \Pi$}
    \For{$j \gets 1, \text{N}$}
        \If{$j \neq k \ \& \ j \neq \text{N}$}
            \State Compute $(u^{\text{min}}_{k,j}, p_{\text{t},k})$ and $(u^{\text{max}}_{k,j}, p_{\text{t},k})$
            \State Determine $[v^{\text{min}}_{k,j}, v^{\text{max}}_{k,j}] = [q^{\text{min}}_{\text{t},k}, q^{\text{max}}_{\text{t},k}] \cap [u^{\text{min}}_{k,j}, u^{\text{max}}_{k,j}]$
            \If{$[v^{\text{min}}_{k,j}, v^{\text{max}}_{k,j}]$ is not empty}
                \State Update the path $\Pi$
                \If{$j \neq 1$}
                    \State Compute $(q^{1}_{\text{t},j}, p_{\text{t},j}) = \text{R}^{-1}_j \circ \text{\textbf{P}}^{-1}_{j,k} (v^{\text{min}}_{k,j}, p_{\text{t},k})$
                    \State Compute $(q^{2}_{\text{t},j}, p_{\text{t},j}) = \text{R}^{-1}_j \circ \text{\textbf{P}}^{-1}_{j,k} (v^{\text{max}}_{k,j}, p_{\text{t},k})$
                    \State $q^{\text{min}}_{\text{t},j} = \min\{q^{1}_{\text{t},j}, q^{2}_{\text{t},j}\}$ and $q^{\text{max}}_{\text{t},j} = \max\{q^{1}_{\text{t},j}, q^{2}_{\text{t},j}\}$
                    \State INTENSITY COMPUTATION($j, q^{\text{min}}_{\text{t},j}, q^{\text{max}}_{\text{t},j}, p_{\text{t},j}, I(p), \Pi$)
                \Else
                    \If{$ k \neq \text{N}$}
                        \State Compute $(q^{1}_{\text{s},1}, p_{\text{s},1}) = \text{\textbf{P}}^{-1}_{1,k} (v^{\text{min}}_{k,1}, p_{\text{t},k})$
                        \State Compute $(q^{2}_{\text{s},1}, p_{\text{s},1}) = \text{\textbf{P}}^{-1}_{1,k} (v^{\text{max}}_{k,1}, p_{\text{t},k})$
                        \State Compute $(q^{1}(\Pi, p), p) = \text{\textbf{M}}_{1,\text{N}}(\Pi)(q^{1}_{\text{s},1}, p_{\text{s},1})$
                        \State Compute $(q^{2}(\Pi, p), p) = \text{\textbf{M}}_{1,\text{N}}(\Pi)(q^{2}_{\text{s},1}, p_{\text{s},1})$
                        \State $q^{\text{min}}(\Pi, p) = \min\{q^{1}, q^{2}\}$ and $q^{\text{max}}(\Pi, p) = \max\{q^{1}, q^{2}\}$
                        \State where $q^{1}=q^{1}(\Pi, p)$ and $q^{2}=q^{2}(\Pi, p)$
                        \State $I(p) = I(p) + q^{\text{max}}(\Pi, p) - q^{\text{min}}(\Pi, p)$
                    \Else
                        \State $I(p) = I(p) + v^{\text{max}}_{k,1} - v^{\text{min}}_{k,1}$
                    \EndIf
                \EndIf
            \EndIf
        \EndIf
    \EndFor
    \EndProcedure
    \end{algorithmic}
    \label{alg:cbrm}
\end{algorithm}

\section{Generalized Algorithm}\label{sec:4}
CBRM only handles parallel beams of light, limiting the algorithm to optical systems consisting of only straight surfaces. We generalize the algorithm to accommodate curved surfaces. Eq. (\ref{eq:12}) defines the intensity in $\text{T}_{\text{N}}$ as the sum of the interval lengths formed by the intersections of the support of the luminance and the line $p=\text{const}$. Therefore, we must still consider a parallel beam of light at the target. The direction of all rays of the beam will vary as it reflects on a curved surface. As a result, the light beam is not necessarily parallel at the other surfaces of the system. In PS this means that position and direction coordinates of all rays in the beam can vary. The beam is no longer represented by a straight line segment on $p=\text{const}$, but by a curved segment for which we generally have no analytical expression.

Intersections in PS cannot be computed analytically since there is no expression for the light beam. We instead discretize the line $p=\text{const}$ in $\text{T}_n$ at the start of the procedure by taking equidistant points and connect them with straight line segments. This discretization is the PS representation of the light beam; when traced to other surfaces it forms a discretized curve $C$.

\begin{figure}[!htb]
    \begin{subfigure}[t]{0.45\textwidth}
        \centering
        \includegraphics[width=\textwidth]{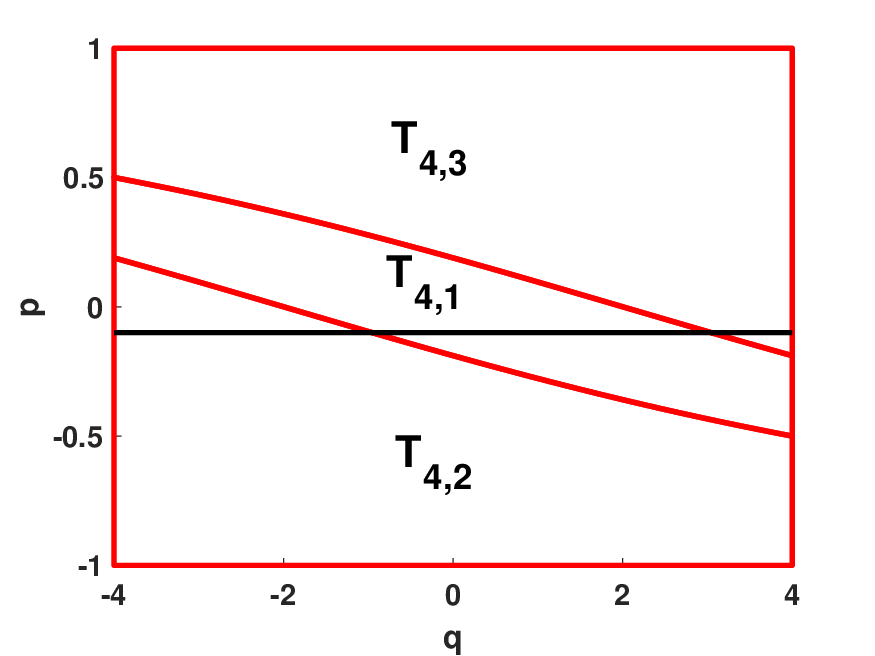}
        \caption{Exact representation of the PS and the line $p=-0.1$.}
        \label{fig:exact}
    \end{subfigure}
    \hspace{0.1\textwidth}
    \begin{subfigure}[t]{0.45\textwidth}
        \centering
        \includegraphics[width=\textwidth]{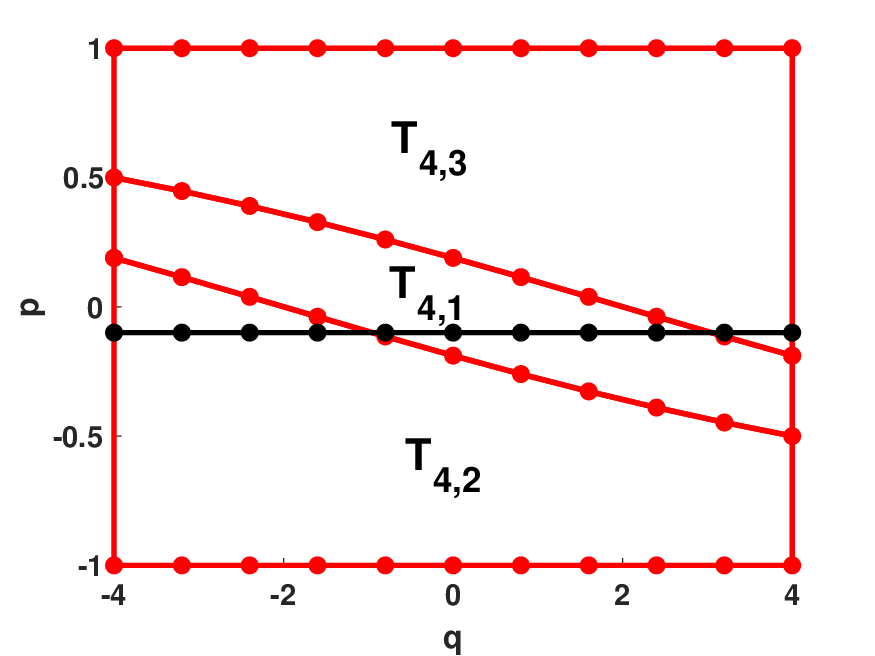}
        \caption{Discretization of the PS and the line $p=-0.1$ used by generalized CBRM.}
        \label{fig:discrete}
    \end{subfigure}
    \caption{The line $p=-0.1$ in $\text{T}_4$ of the CPC.}
    \label{fig:target_cpc}
\end{figure}

In addition to the light beam, the phase spaces of the optical system are also discretized. Recall from Section \ref{sec:2} that there are analytic descriptions of all boundaries of all phase spaces. We discretize each boundary in PS by taking points on these boundaries equidistant in the $q$-direction, and connect them with straight line segments. Fig. \ref{fig:target_cpc} shows $\text{T}_4$ of the CPC with light at $p=-0.1$ and the discretization used by the algorithm. Each discretized PS is stored in a doubly connected edge list (DCEL) \cite{berg2008}. The DCEL stores each region of the PS as a face, each point as a vertex and each line segment as a pair of half-edges. All edges are incident to two faces of the DCEL; therefore, they are split into two half-edges such that each half-edge has exactly one incident face. Each face is bound by the vertices and half edges that form the boundary of the PS region. The half-edges connect pairs of vertices and are ordered counterclockwise around the face they bound. The DCEL is a nice data structure to store geometric information. It makes it easy to perform operations such as traversing the boundary of a given face, accessing a face from an adjacent one if a common edge is given or visiting all edges around a given vertex.

Computing an intersection between $C$ and the boundaries in PS reduces to computing the intersection between two straight line segments; one segment of $C$ and one segment of the discretized boundaries. However, since the PS and the light beam are discretized with many segments we have to check for many pairs of segments if they intersect. To solve this we build a KD-tree \cite{Wald_2006} for each PS. The KD-tree places a bounding box around the PS and subdivides it into increasingly smaller regions storing the boundary segments in its leafs. Non-overlapping regions of PS are stored in the internal nodes and leafs of the KD-tree, i.e., the data structure is a space partition. A boundary segment that intersects different regions of the partition is stored in multiple leafs of the KD-tree since the regions of the KD-tree do not overlap, and they all contain the segment. The tree is a binary tree and every node is split in half along an axis-aligned split plane. We use the surface area heuristic (SAH) \cite{Wald_2006} to select the best splitting plane for every potential split. With the SAH we compute a cost for all possible split planes of a region of the KD-tree. The split plane with the lowest cost is considered to be the best split plane. A region is split when the cost of the best split plane is less than the cost of not splitting the region. The region is not split in half when the cost of the best split plane is greater than the cost of not splitting the region.

Given a segment of $C$ we create a half line by extending the origin of the segment beyond the outer boundaries of PS. The origin of the segment is the endpoint with the smallest $q$-coordinate; it is extended along the direction of the segment in increasing $q$ direction. The segments of the boundaries in PS that can be intersected by the half line are in the cells (leafs) of the KD-tree intersected by the half line; the segments in all cells not intersected by the half line can be ignored. We are interested in the intersection point closest to the origin of the half line. Therefore, we first check for intersections in the cell closest to the origin and continue through the other cells along the half line. When an intersection is found between the half line and a PS segment all subsequent cells of the tree along the ray can be ignored. If the intersection point lies on the segment of $C$ then an intersection between $C$ and a discretized boundary is found; if the intersection point is not on the segment of $C$ then the segment is contained inside a PS region.

The luminance for all possible paths $\Pi$ for a given direction $p \in [-1,1]$ in $\text{T}_n$ is computed recursively, similarly to the original algorithm. The endpoints of the light beam in $\text{T}_n$ are $C^{\min}_{\text{t},\text{N}}$ and $C^{\max}_{\text{t},\text{N}}$. The discretized curve $C_{\text{t},\text{N}}$ intersects various regions in $\text{T}_n$. $C_{\text{N},j}$ is the subset of $C_{\text{t},\text{N}}$ intersecting the region $\text{T}_{\text{N},j}$. The intersection points $C^{\min}_{\text{N},j}$ and $C^{\max}_{\text{N},j}$ with $\partial\text{T}_{\text{N},j}$ are computed analytically. Each subset $C_{\text{N},j}$ corresponds to rays emitted by another surface $j \neq \text{N}$. All segments of $C_{\text{N},j}$ are transformed to a new discretized curve $C_{\text{t},j}$ with endpoints $C^{\min}_{\text{t},j}$ and $C^{\max}_{\text{t},j}$ in $\text{T}_j$. This is done by sequentially applying the maps $\text{\textbf{P}}^{-1}_{j,\text{N}} : \text{T}_{\text{N},j} \to \text{S}_{j,\text{N}}$ and $\text{\textbf{R}}^{-1}_{j} : \text{S}_{j} \to \text{T}_{j}$. The procedure is repeated in $\text{T}_j$. The recursion ends when a subset $C_{k,j}$ is traced back to $\text{S}_1$ or when a subset is empty, like in the original algorithm. If $\text{S}_1$ is found, all points of $C_{\text{s},1}$ are traced to $\text{T}_n$ along $\Pi$ by applying $\text{\textbf{M}}_{1,\text{N}}(\Pi)$. The endpoints $C^{\min}(\Pi, p)$ and $C^{\max}(\Pi, p)$ of $C(\Pi, p)$ are on the boundary of a region $\text{R}_{\text{t}}(\Pi) \subset \text{T}_n$ with positive luminance. The steps of the generalized CBRM algorithm are given in Algorithm \ref{alg:cbrm_ex}.

\subsection{Intensity Distribution of the CPC}\label{sec:4:1}
\begin{figure}[!htb]
    \begin{minipage}{0.45\textwidth}
        \centering
        \includegraphics[width=\textwidth]{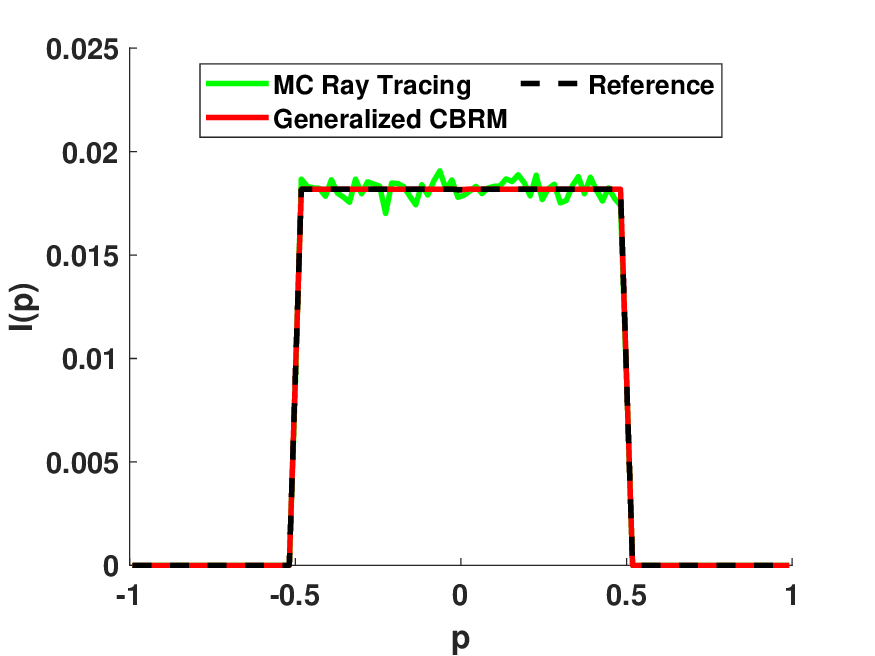}
        \caption{Intensity distributions of the CPC computed with MC ray tracing (green) and generalized CBRM (red) compared to a reference solution.\\}
        \label{fig:intensity_cpc}
    \end{minipage}
    \hspace{0.1\textwidth}
    \begin{minipage}{0.45\textwidth}
        \centering
        \includegraphics[width=\textwidth]{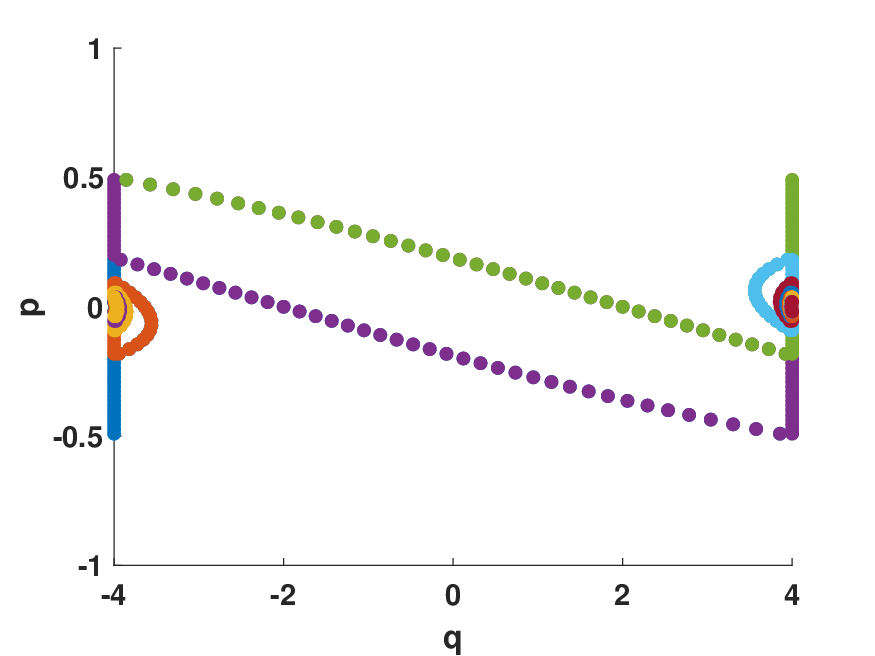}
        \caption{Points on the boundaries of the positive luminance regions in $\text{T}_4$ of the CPC for all paths $\Pi$ with a maximum of 10 reflections. Each color represents a different path.}
        \label{fig:backward_ps_cpc}
    \end{minipage}
\end{figure}

Recall from Section \ref{sec:2:2} that light in the CPC can reflect an infinite number of times, generalized CBRM can run indefinitely as a result. To prevent this we do not consider light that reflects more than 10 times. The reference solution of the CPC is computed with QMC ray tracing using $10^9$ rays that reflect no more than 10 times. The target is divided into 110 bins for (Q)MC ray tracing. The intensity in each bin is also computed with generalized CBRM using Eq. (\ref{eq:13}). The intensity distribution computed with MC ray tracing, shown in Fig. \ref{fig:intensity_cpc}, is again noisy like for the two-faceted cup. Generalized CBRM is able to match the reference solution closely as can be seen in Fig. \ref{fig:intensity_cpc}. MC ray tracing required $10^5$ rays reflecting no more than 10 times, but generalized CBRM required only $1.6 \cdot 10^4$ rays. The light beam and the boundaries in PS were discretized with $10^3$ segments each for the generalized CBRM method. The generalized CBRM algorithm is much more accurate than MC ray tracing and requires far fewer rays to compute the intensity. We can also use generalized CBRM to find the boundaries of the positive luminance regions in $\text{T}_4$; they are shown in Fig. \ref{fig:backward_ps_cpc}.

\begin{algorithm}[H]
    \caption{Recursive procedure to compute the intensity for optical systems containing curved surfaces}
    \textbf{Input}: $k=\text{N}$, $C_{\text{t},k}$ is a discretization of $p=\text{const}$ in $\text{T}_{\text{N}}$, $I(p)=0$, $\Pi = (\text{N})$
    \begin{algorithmic}[1]
        \Procedure{INTENSITY COMPUTATION GENERALIZED}{$k, C_{\text{t},k}, I(p), \Pi$}
        \State Compute $C_{k,j} \ \forall_j \in \{1, \dots, \text{N}-1\}$
        \State $C_{k,j}$ has endpoints $C^{\min}_{k,j}$ and $C^{\max}_{k,j}$
        \For{$j \gets 1, \text{N}-1$}
            \If{$C_{k,j}$ is not empty}
                \State Update the path $\Pi$
                \IIf{$C^{\min}_{k,j} \geq C^{\max}_{k,j}$} inverse the ordering \EndIIf
                \If{$j \neq 1$}
                    \State Compute $C_{\text{t}, j} = \text{\textbf{R}}^{-1}_{j} \circ \text{\textbf{P}}^{-1}_{j,k} (C_{k,j})$
                    \State $C_{\text{t}, j}$ has endpoints $C^{\min}_{\text{t}, j}$ and $C^{\max}_{\text{t}, j}$
                    \IIf{$C^{\min}_{\text{t}, j} \geq C^{\max}_{\text{t}, j}$} inverse the ordering \EndIIf
                    \State INTENSITY COMPUTATION($j, C_{\text{t}, j}, I(p), \Pi$)
                \Else
                    \If{$ k \neq \text{N}$}
                        \State Compute $C_{\text{s},1} = \text{\textbf{P}}^{-1}_{1,k} (C_{k,1})$
                        \State $C_{\text{s},1}$ has endpoints $C^{\min}_{\text{s},1}$ and $C^{\max}_{\text{s},1}$
                        \State Compute $C(\Pi, p) = \text{\textbf{M}}_{1,\text{N}}(\Pi)(C_{\text{s},1})$
                        \State $C(\Pi, p)$ has endpoints $C^{\min}(\Pi, p)$ and $C^{\max}(\Pi, p)$
                        \IIf{$C^{\min}(\Pi, p) \geq C^{\max}(\Pi, p)$} inverse the ordering \EndIIf
                        \State $I(p) = I(p) + C^{\max}(\Pi, p) - C^{\min}(\Pi, p)$
                    \Else
                        \State $I(p) = I(p) + C^{\max}_{k,j} - C^{\min}_{k,j}$
                    \EndIf
                \EndIf
            \EndIf
        \EndFor
        \EndProcedure
    \end{algorithmic}
    \label{alg:cbrm_ex}
\end{algorithm}

\section{Numerical Experiments}\label{sec:5}
We compare generalized CBRM to CBRM and MC ray tracing on the CPC. Recall from Section \ref{sec:2:2} that light in the CPC can reflect an infinite number of times, generalized CBRM can therefore run indefinitely. To prevent this we do not consider light that reflects more than 10 times. This restriction is also applied to CBRM and MC ray tracing for correct comparison. Since CBRM only handles optical systems consisting of straight surfaces, it is applied to a discretized CPC. We discretize each reflector by taking points on the parabola equidistant in the $x$-direction, and connect them with straight line segments. The maximum number of reflections that can occur in the discretized CPC is limited by the number of discrete segments. Recall from \ref{sec:2:2} that a light ray reflects on at most one reflector of the CPC. Therefore, a ray in the discretized CPC can reflect on at most all segments that discretize a reflector. MC ray tracing and generalized CBRM are applied to the regular CPC.

We first compare intensity distributions of the CPC computed with generalized CBRM, CBRM and MC ray tracing to a reference solution. We apply CBRM to a discretized CPC that has 10 segments discretizing each reflector, such that the maximum number of reflections that can occur is equal to the maximum number of reflections we consider. The boundaries in PS and the light beam used by generalized CBRM are all discretized with $10^3$ segments. We compute the reference solution by QMC ray tracing $10^9$ rays that reflect at most 10 times. The target is divided into 110 bins for MC ray tracing. The intensity in each bin is also computed with (generalized) CBRM using Eq. (\ref{eq:13}) where the endpoints of the bin are $p^{m}$ and $p^{m+1}$.

Next, we compare the performance of generalized CBRM for various parameter settings of the algorithm. This is done twice, using 110 and 1010 intervals/bins at the target. We define the performance as the error of the solution compared to the computation time. The error is calculated with:

\begin{equation}\label{eq:error}
    \text{error} = \frac{1}{\text{Ni}} \ \sum_{m=1}^{\text{Ni}}|\hat{I}(p^{m + 1/2}) - \hat{I}_{\text{ref}}(p^{m + 1/2})|,
\end{equation}
where $\text{Ni}$ is the number of bins (intervals) at the target and $\hat{I}_{\text{ref}}$ denotes the reference intensity. We compare five different discretizations of the boundaries in PS using $100 \cdot 2^i$ segments, where $i \in \{0, \dots, 4\}$. We compute the intensity distribution for each PS discretization 10 times using different discretizations of the light beam consisting of $100 \cdot 2^i$ segments, where $i \in \{0, \dots, 9\}$. The reference solution is again computed by QMC ray tracing a number of rays that reflect at most 10 times. We use a reference solution of $10^9$ rays when the target is divided into 110 bins and of $10^{10}$ rays when the target is divided into 1010 bins. The best performing discretization of the boundaries in PS is the one that reaches the smallest error. If the smallest errors are similar then the performance is determined by the time it takes to compute the smallest error. In this case, the best performing discretization of the boundaries in PS is the one with the least computation time.

Finally, we compare the discretizations (of the boundaries in PS) that give the best performance of generalized CBRM for 110 and 1010 bins to the performance of CBRM. The intensity distribution of CBRM is computed using discretizations of $2^i$ segments on the reflectors, where $i \in \{0, \dots, 9\}$. The performance of CBRM is also defined as the computation time compared to the error of the solution given in Eq. (\ref{eq:error}). We compute the error of CBRM using the reference solutions from the previous experiment.

\section{Results}\label{sec:6}
The aim of the research was to introduce a generalization to the concatenated backward ray mapping algorithm. We discussed the original backward method and our generalized method. Both algorithms were used to compute the intensity distribution on the compound parabolic concentrator.

\begin{figure}[htb]
    \begin{subfigure}{0.45\textwidth}
        \centering
        \includegraphics[width=\textwidth]{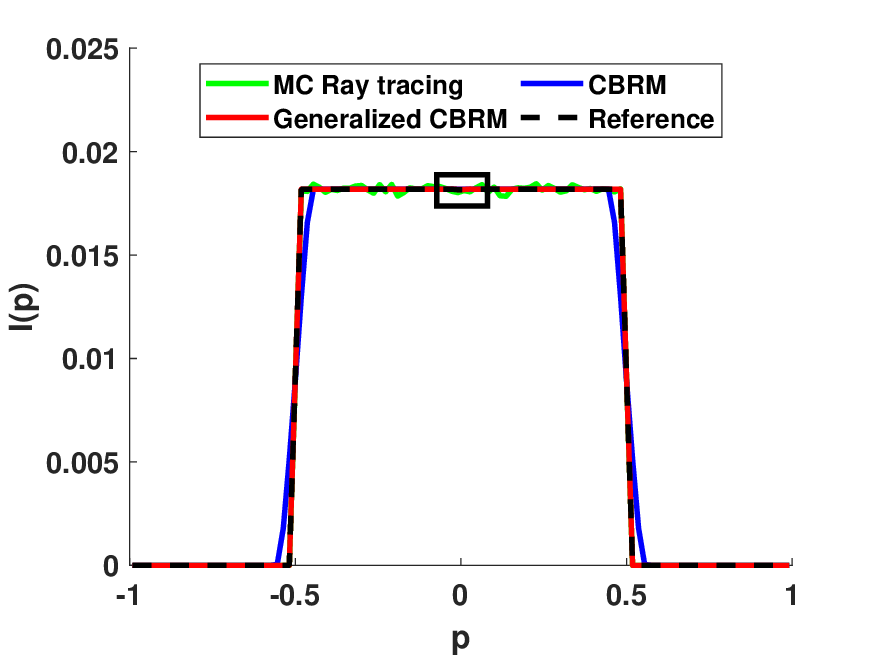}
        \caption{Comparison of intensity distributions computed for the CPC.}
        \label{fig:comparison}
    \end{subfigure}
    \hspace{0.1\textwidth}
    \begin{subfigure}{0.45\textwidth}
        \centering
        \includegraphics[width=\textwidth]{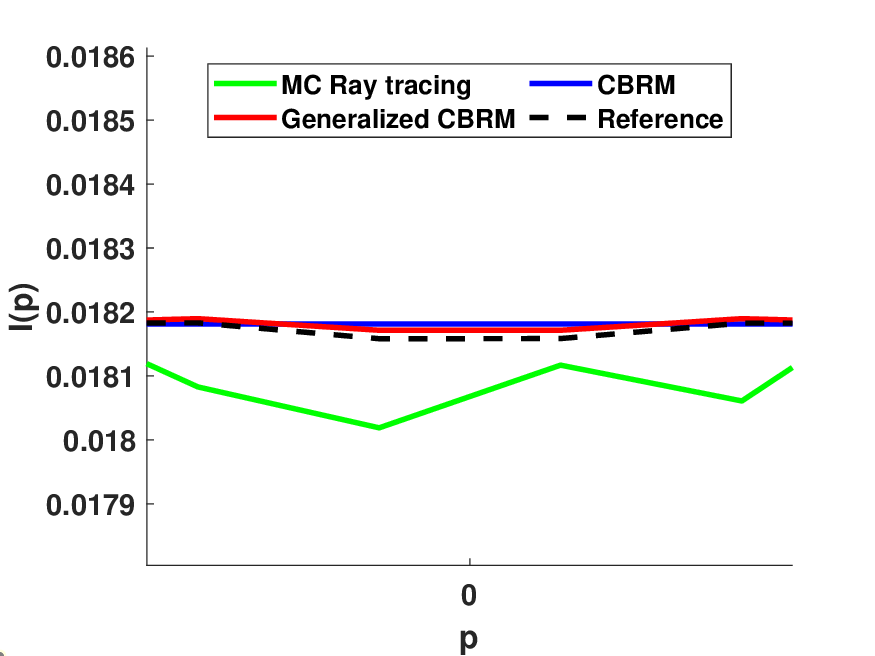}
        \caption{Zoom of the indicated area of the intensity distributions.}
        \label{fig:comparison_zoom}
    \end{subfigure}
    \caption{Comparison of intensity distributions of the CPC. We compare MC ray tracing (green), CBRM (blue) and generalized CBRM (red) to a reference solution computed with QMC ray tracing.}
    \label{fig:comparison_parent}
\end{figure}

Fig. \ref{fig:comparison} and Fig. \ref{fig:comparison_zoom} show three intensity distributions of the CPC computed by generalized CBRM, CBRM and MC ray tracing compared to a reference solution. The figures clearly show the differences between the intensity distributions found by the algorithms. MC ray tracing computed a noisy intensity distribution with a profile similar to that of the reference solution. CBRM on the other hand found an intensity distribution with a profile that differs from the profile of reference solution. Finally, generalized CBRM found an intensity distribution closely matching the profile of the reference solution and without any noise. It took $3.6 \cdot 10^4$ rays to compute the CBRM solution, $1.6 \cdot 10^4$ rays to compute the generalized CBRM solution and $10^6$ rays to compute the MC ray tracing solution. MC ray tracing gives an intensity distribution with steep sides that is similar to the reference solution because it was applied to the regular CPC. However, the intensity distribution is noisy due to the random selection of rays at the source. This is in agreement with the noisy distributions of the cup and the CPC that were previously discussed in section 3.1 (Fig. \ref{fig:intensity_cup}) and section 4.1 (Fig. \ref{fig:intensity_cpc}). CBRM \cite{Filosa2021} on the other hand was applied to a discretization of the CPC where each reflector is replaced by 10 straight line segments. This optical system is different from the CPC, so it also behaves differently. As a result CBRM gave an intensity distribution without steep sides that differs from the reference solution. A finer discretization of the CPC makes the systems more alike, so the distributions should also be more alike. Generalized CBRM did not require the optical system to be discretized, and it did not use a random set of light rays. It computed an intensity distribution with steep sides that is similar to the reference solution without any noise. What stands out in these results is that generalized CBRM is more accurate than MC ray tracing and CBRM but required fewer rays to compute the intensity distribution.

\begin{figure}[htb]
    \begin{subfigure}{0.45\textwidth}
        \centering
        \includegraphics[width=\textwidth]{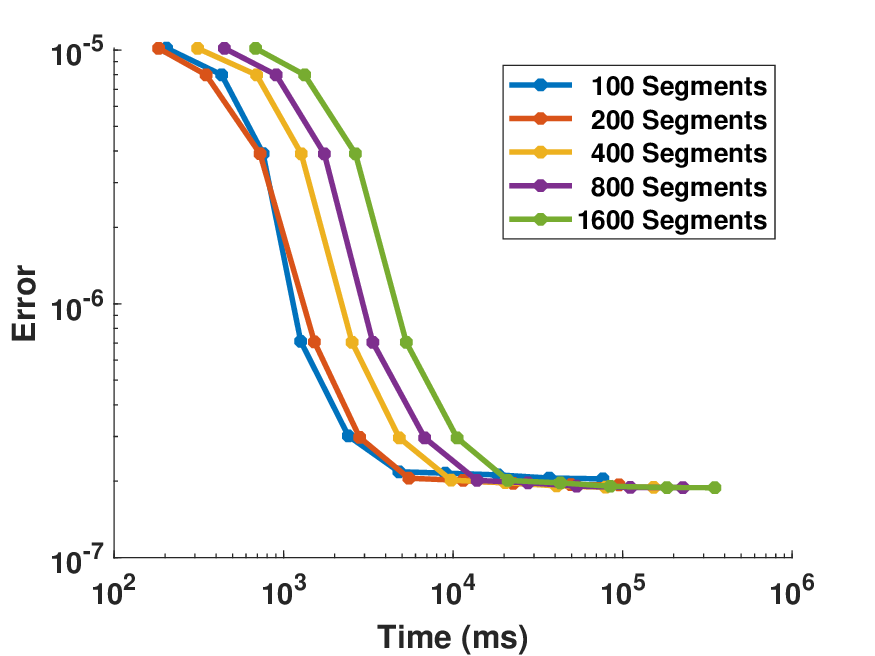}
        \caption{CBRM performance for 110 bins at the target.}
        \label{fig:discretization_error_110}
    \end{subfigure}
    \hspace{0.1\textwidth}
    \begin{subfigure}{0.45\textwidth}
        \centering
        \includegraphics[width=\textwidth]{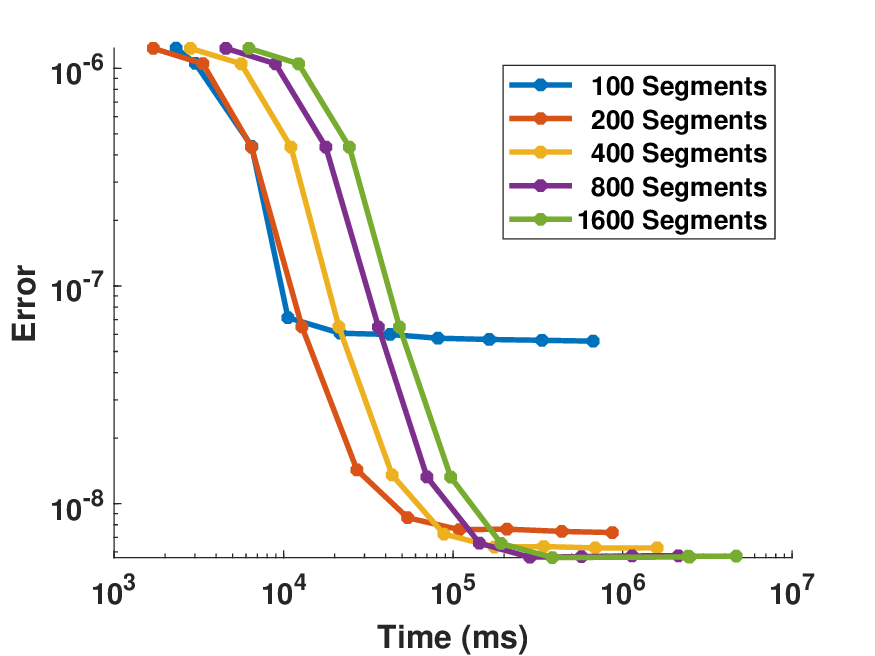}
        \caption{CBRM performance for 1010 bins at the target.}
        \label{fig:discretization_error_1010}
    \end{subfigure}
    \caption{Performance of generalized CBRM on the CPC. The error is the difference between the intensity distribution computed with generalized CBRM and a reference solution computed with QMC ray tracing. We compare PS discretizations of 100 (blue), 200 (orange), 400 (yellow), 800 (purple) and $1.6 \cdot 10^3$ (green) segments.}
    \label{fig:discretization_error}
\end{figure}

Fig. \ref{fig:discretization_error_110} compares the performance of generalized CBRM for different parameter settings of the algorithm using 110 intervals/bins at the target surface. Every curve was computed using a different discretization of the boundaries in PS. What is interesting about Fig. \ref{fig:discretization_error_110} that all curves have a similar shape, which implies that using a finer discretizations of the boundaries has only a small effect on the performance of generalized CBRM. The points of each curve were computed using different discretizations of the light beam. The error of all curves became much smaller as the number of light beam segments increased. We can also see in Fig. \ref{fig:discretization_error_110} that all curves stopped improving at the same discretization of the light beam. The discretization of the light beam had a larger effect on performance than the discretization of the boundaries in PS because of the recursion that happens in generalized CBRM. During every iteration, generalized CBRM computes the intersection between the light beam and the regions in PS; the algorithm continues for each intersection segment. As a result, generalized CBRM computes using a smaller subset of the discretized light beam after each reflection. Therefore, it is important to discretize the light beam with enough segments to ensure that it still closely resembles the light beam after many reflections. Still there is a point after which increasing the number of discrete segments on the light beam no longer improves the accuracy of the solution.

The test of Fig. \ref{fig:discretization_error_110} was repeated in Fig. \ref{fig:discretization_error_1010} this time using 1010 bins at the target. The results are similar to those of the previous test, but there are two notable differences. The algorithm was more accurate because more bins were used at the target which allowed it to use more PS information. Furthermore, using only 100 segments to discretize the boundaries in PS significantly reduced the performance of the algorithm. This happened because generalized CBRM used more PS information to compute the intensity profile. As a result it also required a better discretization of the boundaries in PS. This shows that there is a correlation between the number of bins at the target and the minimum discretization of the boundaries in PS required by generalized CBRM.

\begin{figure}[htb]
    \begin{subfigure}{0.45\textwidth}
        \centering
        \includegraphics[width=\textwidth]{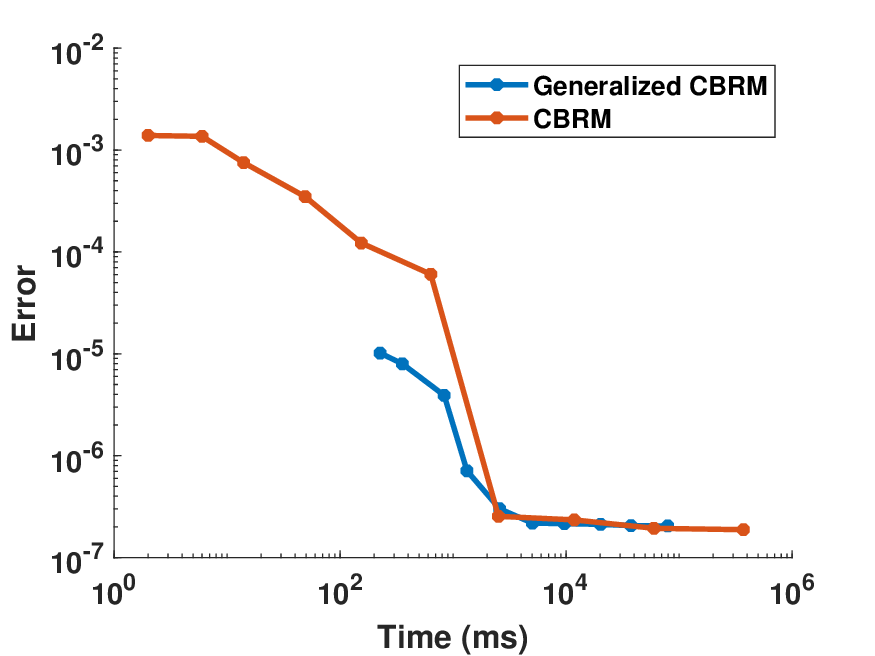}
        \caption{Performance of CBRM compared to generalized CBRM for 110 bins at the target.}
        \label{fig:error_comparison_110}
    \end{subfigure}
    \hspace{0.1\textwidth}
    \begin{subfigure}{0.45\textwidth}
        \centering
        \includegraphics[width=\textwidth]{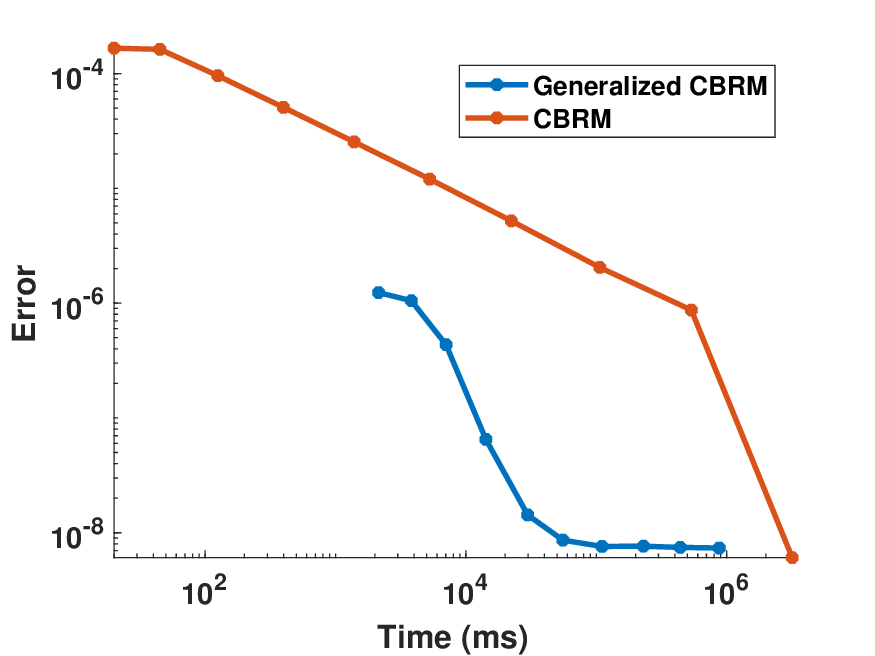}
        \caption{Performance of CBRM compared to generalized CBRM for 1010 bins at the target.}
        \label{fig:error_comparison_1010}
    \end{subfigure}
    \caption{Performance of generalized CBRM (blue) compared to CBRM (orange) on the CPC. The error is the difference between the intensity distributions and a reference solution computed with QMC ray tracing.}
    \label{fig:error_comparison}
\end{figure}

We compared generalized CBRM, using the PS discretization with the best performance, to the performance of CBRM. In Fig. \ref{fig:discretization_error_110}, all discretizations of the boundaries in PS reach a similar error value. So, the best result for the case of 110 bins at the target is the discretization with 100 boundary segments since it takes the least computation time. In Fig. \ref{fig:discretization_error_1010}, the discretizations also reach a similar error except for the discretization of 100 boundary segments. So, the best result for the case of 1010 bins at the target is the discretization with 200 boundary segments. The comparison of CBRM and generalized CBRM for the case of 110 bins at the target is shown in Fig. \ref{fig:error_comparison_110}. The most striking observation in this figure is the sudden decrease of the CBRM error for a discretization of 64 reflector segments. The figure also shows that the generalized CBRM error is smaller than the CBRM error in most cases but that CBRM reaches maximum precision slightly faster than generalized CBRM. The sudden decrease of the CBRM error can be explained by the target PS of the target of the discretized CPC. The discretized CPC is different from the CPC meaning that the non-empty regions in the target PS of the target are also different from those of the CPC. Using a discretization with more reflector segments makes the optical systems and the non-empty regions in the target PS of the target more alike. At a certain discretization the shape of the non-empty regions is so similar to those of the CPC that it resulted in a sudden increase in accuracy.

Fig. \ref{fig:error_comparison_1010} compares the performance of CBRM and generalized CBRM for the case where there are 1010 bins at the target. What stands out in this figure is the performance difference between CBRM and generalized CBRM. Generalized CBRM behaved similarly to the case of 110 bins at the target. However, the number of bins at the target had a large effect on the performance of CBRM. It took a much finer discretization of the CPC to reach maximum precision. When CBRM computes with more bins it uses more information about the target PS of the target. As a consequence the algorithm requires a finer discretization of the CPC to make the shape of the non-empty regions so similar to those of the CPC that it results in a sudden increase in accuracy.

\section{conclusion}\label{sec:7}
In this paper we introduced a generalized concatenated backward ray mapping (CBRM) algorithm that handles optical systems with curved surfaces. We presented a more general description of the boundaries in phase space (PS), allowing us to compute the source and target PS of curved optical surfaces. We modified the original CBRM algorithm to be able to compute the intensity distribution for optical systems containing curved surfaces. The generalized algorithm uses a discretization of the phase spaces of an optical system and a discretization of the PS representation of the light beam. We implemented a doubly connected edge list (DCEL) and KD-tree data structure to compute more efficiently in PS. CBRM and generalized CBRM are applied to the compound parabolic concentrator (CPC), a standard optical system that reshapes light from a Lambertian source into a focused beam.

MC ray tracing is limited by the random selection of rays at the source because of which the intensity distribution is noisy. The number of bins at the target has a large effect on CBRM. The CPC must be discretized by many reflection segments if the number of bins is large. Otherwise, the CBRM is not accurate enough. However, if the CPC is discretized with many segments CBRM is slow. Generalized CBRM on the other hand is effected by the discretization of the light beam and the discretization of the boundaries in PS. The number of bins at the target has no effect on generalized CBRM. Of the two discretizations the discretization of the light beam has the biggest effect on performance. The number of light beam segments required for the discretization depends on the maximum number of reflections for which is the algorithm is computed; more segments are required for more reflections. Generalized CBRM computes equally good or better compared to CBRM depending on the number of bins at the target.

We have shown that there are general expressions of the boundaries in PS and that they can be used to compute the phase spaces of an optical system with curved surfaces. We introduced a generalized backward ray mapping algorithm that uses this PS information to compute the intensity distribution of an optical system. Our results showed that generalized CBRM computes the intensity distribution of the CPC faster and more accurately than the original CBRM method and Monte Carlo ray tracing.

Future research may involve exploring splines instead of straight segments to discretize the light beam and the boundaries in PS which could have an interesting effect on performance. It is also interesting to add refraction to the algorithm. This would allow us to compute the intensity distribution of many more 2D optical systems. Another research direction is to extend the method to 3D optical systems which will make it more applicable; a first step could be to find a suitable representation of the boundaries in PS for 3D optical systems.

\bmhead{Abbreviations}
MC, Monte Carlo; QMC, Quasi Monte Carlo; PS, phase space; CBRM, concatenated backward ray mapping; DCEL, doubly connected edge list; CPC, compound parabolic concentrator.

\bmhead{Acknowledgments}
Not applicable.

\section*{Declarations}               
\bmhead{Funding}
This work was funded by NWO and Technische Wetenschappen in the framework of the project "3D phase space ray tracing for the illumination optics industry", project 17959.
               
\bmhead{Competing interest}
The authors declare that they have no competing interests.

\bmhead{Ethics approval} 
Not applicable.
                              
\bmhead{Consent to participate}
Not applicable.

\bmhead{Consent for publication}
Not applicable.

\bmhead{Availability of data and materials}
Please contact the corresponding author for data requests.
                              
\bmhead{Code availability}
The simulation code for obtaining the results presented in this paper is not publicly available at this time but may be obtained from the authors upon request.

\bmhead{Authors' contributions}
All authors are equal contributors to this paper. All authors read and approved the final manuscript.

\bibliography{bibliography}
\end{document}